\documentstyle[12pt]{article}

\catcode`\@=11
\def\marginnote#1{}
\newcount\hour
\newcount\minute
\newtoks\amorpm
\hour=\time\divide\hour by60
\minute=\time{\multiply\hour by60 \global\advance\minute by-\hour}
\edef\standardtime{{\ifnum\hour<12 \global\amorpm={am}%
        \else\global\amorpm={pm}\advance\hour by-12 \fi
        \ifnum\hour=0 \hour=12 \fi
        \number\hour:\ifnum\minute<10 0\fi\number\minute\the\amorpm}}
\edef\militarytime{\number\hour:\ifnum\minute<10 0\fi\number\minute}
\def\draftlabel#1{{\@bsphack\if@filesw {\let\thepage\relax
   \xdef\@gtempa{\write\@auxout{\string
      \newlabel{#1}{{\@currentlabel}{\thepage}}}}}\@gtempa
   \if@nobreak \ifvmode\nobreak\fi\fi\fi\@esphack}
        \gdef\@eqnlabel{#1}}
\def\@eqnlabel{}
\def\@vacuum{}
\def\draftmarginnote#1{\marginpar{\raggedright\scriptsize\tt#1}}
\def\draft{\oddsidemargin -.5truein
        \def\@oddfoot{\sl preliminary draft \hfil
        \rm\thepage\hfil\sl\today\quad\militarytime}
        \let\@evenfoot\@oddfoot \overfullrule 3pt
        \let\label=\draftlabel
        \let\marginnote=\draftmarginnote
   \def\@eqnnum{(\theequation)\rlap{\kern\marginparsep\tt\@eqnlabel}%
\global\let\@eqnlabel\@vacuum}  }

\def\preprint{\twocolumn\sloppy\flushbottom\parindent 1em
        \leftmargini 2em\leftmarginv .5em\leftmarginvi .5em
        \oddsidemargin -.5in    \evensidemargin -.5in
        \columnsep 15mm \footheight 0pt
        \textwidth 250mmin      \topmargin  -.4in
        \headheight 12pt \topskip .4in
        \textheight 175mm
        \footskip 0pt
        \def\@oddhead{\thepage\hfil\addtocounter{page}{1}\thepage}
        \let\@evenhead\@oddhead \def\@oddfoot{} \def\@evenfoot{} }

\def\titlepage{\@restonecolfalse\if@twocolumn\@restonecoltrue\onecolumn
     \else \newpage \fi \thispagestyle{empty}\c@page\z@ 
        \def\thefootnote{\fnsymbol{footnote}} }

\def\endtitlepage{\if@restonecol\twocolumn \else  \fi
        \def\thefootnote{\arabic{footnote}}
        \setcounter{footnote}{0}}  

\catcode`@=12
\relax
\def\beq{\begin{equation}}
\def\eeq{\end{equation}}

\def\Im{\mathop{\rm Im}}
\def\NP#1#2#3{Nucl. Phys. \underline{#1} (19#2) #3}
\def\ov{\overline}
\def\PL#1#2#3{Phys. Lett. \underline{#1} (19#2) #3}
\def\PR#1#2#3{Phys. Rev. \underline{#1} (19#2) #3}
\def\PRL#1#2#3{Phys. Rev. Lett. \underline{#1} (19#2) #3}
\def\Re{\mathop{\rm Re}}

\def\crbig{\\\noalign{\vspace {3mm}}}
\def\bigint{{\displaystyle\int}}
\def\Dint{{\int d^2\theta d^2\ov\theta\,}}
\def\Fint{{\int d^2\theta\,}}
\def\Fbarint{{\int d^2\ov\theta\,}}

\def\Btwo{{{\cal B}_2}}
\def\B{{\cal B}}
\def\C{C_3}
\def\Csix{C_6}
\def\Hthree{H_3}
\def\G{G_4}
\def\Gseven{G_7}

\def\Dfive{{\Delta_5}}
\def\Dfour{{\Delta_4}}

\def\Mele{{{\cal M}_{11}}}
\def\Mten{{{\cal M}_{10}}}
\def\Wsix{{{\cal W}_6}}
\def\Mfour{{{\cal M}_4}}
\def\Ztwo{{{\bf Z}_2}}
\def\Ctwo{{{\cal C}_2}}

\def\L{{\cal L}}
\def\S{S}

\relax

\begin{document}
\topmargin-2.4cm
\renewcommand{\theequation}{\thesection.\arabic{equation}}
%
\begin{titlepage}
\begin{flushright}
NEIP--00-017 \\
hep--th/0009054 \\
September 2000
\end{flushright}
\vspace{2.4cm}

\begin{center}
{\Large\bf A Five-brane Modulus in the Effective \\[3mm]
N=1 Supergravity of M-Theory$^*$}
\vskip .8in
{\bf J.-P. Derendinger and R. Sauser}$^{\dagger}$
\vskip .1in
{\it Institute of Physics,
University of Neuch\^atel \\
CH--2000 Neuch\^atel, Switzerland}
\end{center}

\vspace{1cm}

\begin{center}
{\bf Abstract}
\end{center}
\begin{quote}
Five-branes lead in four dimensions to massless $N=1$ supermultiplets 
if M-theory is compactified on $S^1/\Ztwo\times$ (a Calabi-Yau threefold). 
One of them describes the 
modulus associated with the position of the five-brane along the circle 
$S^1$. We derive the effective four-dimensional supergravity of this 
multiplet and its coupling to bulk moduli and to Yang-Mills and charged 
matter multiplets located on $\Ztwo$ fixed planes. The
dynamics of the five-brane modes is obtained by reduction and 
supersymmetrization of the covariant five-brane bosonic action.
Our construction respects all symmetries of M-theory, including the 
self-duality of the brane antisymmetric tensor. Corrections to gauge 
couplings are strongly constrained by this self-duality property.
The brane contribution to the effective scalar potential is formally 
similar to a renormalization 
of the dilaton. The vacuum structure is not modified. 
Altogether, the impact of the five-brane modulus 
on the effective supergravity is reminiscent of string one-loop 
corrections produced by standard compactification moduli. 
\end{quote}

\vfill
\begin{flushleft}
\rule{8.1cm}{0.2mm}\\
$^{\star}$
{\small Research supported in part by
the Swiss National Science Foundation.} \\
$^{\dagger}$ {\small\tt jean-pierre.derendinger, 
roger.sauser@unine.ch}
\end{flushleft}

\end{titlepage}
\setcounter{footnote}{0}
\setcounter{page}{0}
\setlength{\baselineskip}{.7cm}
\newpage 
\newpage

\section{Introduction}\label{secintro}
\setcounter{equation}{0}

Compactification of M-theory on
\beq\label{O7is}
O_7 = X_6\times S^1/\Ztwo\,,
\eeq
with a Calabi-Yau threefold $X_6$, leads to $N=1$ 
supersymmetry in four space-time dimensions. 
Five-branes configurations preserve this supersymmetry
if their world-volume $\Wsix$ is suitably aligned: it should 
enclose four-dimensional Minkowski space $\Mfour$ and a holomorphic
two-cycle ${\cal C}_2$ in $X_6$ \cite{BeckerBS, BVS, WITTENSTRONG}:
$\Wsix=\Mfour\times {\cal C}_2$.
With this embedding, five-brane massless excitations on the world-volume,
which belong to a tensor multiplet of chiral six-dimensional
supersymmetry on $\Wsix$ \cite{GT,KM}, produce in the low-energy 
four-dimensional effective supergravity various $N=1$ 
multiplets of massless fields. Some of these modes 
are deeply related to the Calabi-Yau geometry, 
and computing their effective theory is a very complicated task.
There are however universal
modes which can be more easily described, the most obvious example 
being the real scalar associated to the position of the five-brane on the
orbifold $S^1/\Ztwo$. This `universal five-brane modulus' will be
the main subject of the present paper: we will compute its
effective supergravity couplings to the modes of M-theory on $O_7$
which are also perturbative massless states of $E_8\times E_8$
heterotic strings on $X_6$. In the simplest case of the standard 
embedding, these are the $N=1$ supergravity and dilaton multiplets, 
the modulus of the Calabi-Yau volume, $E_6\times E_8$ gauge fields 
and chiral matter in representation $({\bf 27}, {\bf 1})$.
Lukas, Ovrut and Waldram (LOW) \cite{LOW} have derived
the effective supergravity for these heterotic states in 
a non-trivial background value of the brane modulus. 
Our goal here is to obtain a complete\footnote{Including terms with
two derivatives or less.} effective supergravity for 
the supermultiplet of the universal brane modulus. 

When computing an effective Lagrangian, it is usually important
to respect the symmetries of the underlying theory. For instance,
the tensor multiplet of five-brane excitations has an antisymmetric tensor
with a self-dual field strength. This symmetry has important 
implications in four dimensions: the effective theory has a 
massless antisymmetric tensor dual to a pseudoscalar or, in terms
of supermultiplets a chiral multiplet dual to a linear 
multiplet\footnote{This property is called `chiral-linear duality' 
in the paper.} \cite{Sie}. This observation has immediate implications
on the effective supergravity of the brane modes since only
a limited class of chiral multiplets couplings is allowed by chiral-linear 
duality \cite{FGKVP}. Another example is the 
fact that M-theory on $O_7$ can be
defined by specific Bianchi identities. Their symmetry properties  
provide information on the supergravity multiplets to be used in their 
effective description. In ref. \cite{DS}, we have formulated the effective 
supergravity of M-theory on $O_7$ without five-branes using 
Lagrange multiplier superfields to impose, by their field equations, 
these Bianchi identities and all their symmetries. This formulation 
is well adapted to the inclusion of five-brane modes.

The construction reveals some interesting features. The contributions
of the five-brane universal modulus are closely similar to the 
perturbative corrections generated by volume moduli \cite{DKL, DFKZ,
CO}. In particular, 
gauge threshold corrections arise, with a gauge-group independent term 
linked by supersymmetry to brane kinetic terms. This correction can
be regarded as a renormalization of the dilaton field. Of course,
there is no induced superpotential and the vacuum properties of the
scalar potential are not severely modified. The physics impact of
the five-brane fields is in the modification of the M-theory background
equation (the `cohomology condition' \cite{WITTENSTRONG}) and  
in the gauge-group-dependent threshold corrections, as observed by LOW
\cite{LOW}.

The present article is divided in three parts. In section 2, we
study the role of the six-form field which couples naturally to 
the five-brane. We construct a version of the bosonic sector of 
eleven-dimensional supergravity in which the field equation 
of the six-form field is the required Bianchi identity. This theory 
can then easily be coupled to contributions arising from $S^1/\Ztwo$ 
fixed planes or from five-branes. Its reduction on $O_7$ provides the 
link with the effective four-dimensional supergravity derived in ref. 
\cite{DS} and a guiding line for the introduction of five-brane fields. 
The section ends with a first glance at the Calabi-Yau background 
equation.

Section 3 is devoted to the dynamics of the five-brane massless modes. 
Our starting point is the self-dual formulation of the bosonic five-brane 
action, with an auxiliary scalar, as derived by Pasti, Sorokin and Tonin (PST) 
\cite{5bPST}. The $O_7$ truncation is performed and supersymmetrized, in flat 
space and in an eleven-dimensional supergravity background. 
The resulting kinetic Lagrangian for the five-brane modulus multiplet
possesses as expected chiral-linear duality: the brane
modulus can be either described 
by a linear supermultiplet $\hat L$ or by a chiral $\hat S$ with
symmetry $\hat S \longrightarrow\hat S +ic$ ($c$ a real constant).
This invariance severely restricts the possible form of the brane 
contributions in the Lagrangian. 
We also discuss how the various contributions to the 
scalar potential cancel each other. 

The complete effective supergravity coupled to the five-brane Lagrangian 
is the subject of section 4. Following the procedure valid for the Calabi-Yau
volume modulus $T$, we introduce threshold corrections as the most 
general term allowed by the shift symmetry acting on the brane multiplet 
$\hat S$. We then consider the two dual versions of the effective 
supergravity, with the dilaton embedded either in a chiral or in a 
linear multiplet. The analysis of the gauge couplings in the linear
version reveals a universal quadratic correction generated by the brane 
kinetic terms, and a linear dependence in the threshold terms. In the 
chiral version of the theory, the quadratic correction is moved into 
the K\"ahler potential of the chiral dilaton $S+\ov S$. This result is 
strongly similar to standard gauge threshold corrections in the modulus 
$T$, which are perturbative one-loop contributions in string theory. 
We then compare our expressions with the background found by
LOW and discuss the modifications of the scalar potential introduced 
by the brane modulus. 

Finally, section 5 contains some concluding remarks and
an appendix defines our conventions\footnote{
We use the same notations as in ref. \cite{DS} and our
conventions for supergravity expressions are (mostly) as in ref. 
\cite{KU}.}.

\section{The six-form field}\label{secsixform}
\setcounter{equation}{0}

In this section, we first discuss a formulation of the bosonic sector of
eleven-dimensional supergravity in which the Bianchi identity for 
the four-form $\G$ is explicitly given by the field equation of a 
six-form $\Csix$. This eleven-dimensional field plays the role of a 
multiplier and its Lagrangian can be easily modified to include
source contributions arising, for instance, from five-branes\footnote{
Our procedure is similar, but not identical, to the method of 
de Alwis \cite{deA}}. 
Explicitly, the modified Lagrangian is of the form 
$\Csix\wedge(d\G-\Delta_5)$, where $\Delta_5$ is the
five-form source of the Bianchi identity. It also turns out to be
at the origin of the four-dimensional `Lagrange multiplets' 
described in a previous publication \cite{DS}, in which Bianchi
identities were field equations. 
After having introduced $\Csix$ at the level of the 
bosonic sector of the standard Cremmer-Julia-Scherk  
eleven-dimensional supergravity \cite{CJS}, we 
consider the modifications required by the two ten-dimensional  
planes fixed under $\Ztwo$ and by the presence of five-branes. 

\subsection{Eleven-dimensional supergravity}\label{secsixform1}

We begin by considering the standard 
CJS formulation \cite{CJS}. In terms of differential forms, 
the bosonic part of the eleven-dimensional supergravity action is given by
\beq\label{SCJS}
2\kappa_{11}^2 S_{\rm CJS}=
-\int_{\Mele}{eR} 
-{1\over 2}\int_{\Mele}{\G\wedge *\G}
-{1\over 6}\int_{\Mele}{\C\wedge \G\wedge\G},
\eeq
where the two independent fields are the metric 
(vielbein) and the three-form potential $\C$. The four-form field 
strength $\G$ is defined by $\G=d\C$, and $\Mele$ is 
eleven-dimensional Minkowski space. 
The equation of motion for $\C$ that can be computed from the 
action (\ref{SCJS}) is
\beq\label{standardeom}
\C \,:\,\, d*\!\G=-{1\over 2}\G\wedge\G,
\eeq
and the Bianchi identity reads $d\G =0$.
Note that $S_{\rm CJS}$ is invariant under the standard 
gauge transformation 
\beq\label{Mgaugetransf}
\C \rightarrow \C + d\Lambda_2,
\eeq
where $\Lambda_2$ is a two-form.

Since we would like to incorporate ``magnetic'' five-branes\footnote{
As opposed to the ``electric'' membranes which naturally couple to 
the CJS action (\ref{SCJS}).}
in our discussion, it is 
natural to look for an action 
which contains a seven-form field strength $\Gseven$ dual to the 
usual four-form $\G$. The 
structure of the topological term $\C\wedge\G\wedge\G$ 
in (\ref{SCJS}) does not 
allow us to completely eliminate the three-form $\C$, and an action 
trivially equivalent to $S_{\rm CJS}$ is
\beq\label{S11dsd}
\begin{array}{rcl}
2\kappa_{11}^2 S_{11 \rm sd}&=&
-\int_{\Mele}{eR} 
-{1\over 2}\int_{\Mele}{\G\wedge *\G}
-{1\over 6}\int_{\Mele}{\C\wedge d\C\wedge d\C} \crbig
&&+\int_{\Mele}{\Gseven\wedge (\G-d\C)},
\end{array}
\eeq
where the four independent fields are now the metric (vielbein), 
the three-form $\C$, the four-form $\G$ and the seven-form $\Gseven$.
The equations of motion for the antisymmetric tensor fields are
\beq
\begin{array}{rclrcl}
\C&:&\hfill d\Gseven&=&-{1\over 2}d\C\wedge d\C, \hfill
\crbig
\G&:&\hfill *\G&=&\Gseven, \hfill
\crbig
\Gseven&:&\hfill \G&=&d\C. \hfill
\end{array}
\eeq
They are certainly equivalent to the original 
field equation in the CJS version of the theory. 
The solution of the equation for $\C$ is
\beq\label{G7}
\Gseven= -d\Csix-{1\over 2}\C\wedge d\C,
\eeq
where $\Csix$ is an arbitrary six-form potential. 
Notice that the invariance of the seven-form $\Gseven$ 
under the gauge transformation (\ref{Mgaugetransf}) imposes that
\beq\label{C6gaugetransf}
\Csix \rightarrow \Csix - {1\over 2}\Lambda_2\wedge d\C + d\Lambda_5.
\eeq
We can now write a more interesting form of the bosonic sector of 
eleven-dimensional supergravity in which the Bianchi identity is imposed 
via by six-form field $\Csix$.
Substituting the expression (\ref{G7}) for $\Gseven$ into 
the action (\ref{S11dsd}),
we obtain (with a partial integration) a formulation where the 
four independent fields are the metric, $\C$, $\G$ and $\Csix$:
\beq\label{S11dsd'}
\begin{array}{rcl}
2\kappa_{11}^2 S_{11\rm sd'}&=&
-\int_{\Mele}{eR} 
-{1\over 2}\int_{\Mele}{\G\wedge *\G}
-{1\over 2}\int_{\Mele}{\C\wedge d\C\wedge (\G-{2\over 3}d\C)}
\crbig
&& +\int_{\Mele}{\Csix\wedge d\G}.
\end{array}
\eeq
This action is invariant under gauge symmetries (\ref{Mgaugetransf}) 
and (\ref{C6gaugetransf}).
The equations of motion for $\G$, $\Csix$ and $\C$ are now
\beq
\begin{array}{rclrcl}
\G&:&\hfill *\G&=& -d\Csix-{1\over 2}\C\wedge d\C\,, \hfill
\crbig
\Csix&:&\hfill d\G&=&0 \,, \hfill
\crbig
\C&:&d\C\wedge (d\C-\G)
&=&-{1\over 2}\C\wedge d\G \,. \hfill
\end{array}
\eeq
The exterior derivative of the first equation is the CJS 
equation (\ref{standardeom}) if in addition $\G=d\C$. 
The second relation is the Bianchi identity 
which says that locally $\G=dA_3$. Finally, the third equation 
implies that $\C$ and $A_3$ can differ by a gauge transformation 
(\ref{Mgaugetransf}) and by irrelevant particular solution to 
eq. (\ref{standardeom}).

The $O_7$ truncation of theory (\ref{S11dsd'}) is as follows. 
Under $\Ztwo$, $\C$ and $\G$ are as usual odd while $\Csix$ is even. 
Since $\Ztwo$ acts on the $S^1$ coordinate $x^4$,
the universal massless modes of $\Csix$ surviving the truncation 
will be $C_{\mu\nu\rho\sigma i\ov j}$, $C_{\mu\nu ij\ov i\ov j}$,
$C_{ijk\ov i\ov j\ov k}$ and $C_{\mu\nu\rho ijk}$ (as well as the conjugate
$C_{\mu\nu\rho\ov i\ov j\ov k} = C_{\mu\nu\rho ijk}^*$). Their field
equations respectively imply the Bianchi identity for
$(d\G)_{4 ij\ov i\ov j}$ which is a background equation since it is 
not associated to any four-dimensional massless mode,
and the four-dimensional Bianchi identities
for the massless components $G_{4\rho k\ov k}$, $G_{4\mu\nu\rho}$
and $G_{4\ov i\ov j\ov k}$. Notice also that the modified
topological term in action (\ref{S11dsd'}) is eliminated by the truncation.
The resulting truncated four-dimensional action is trivial 
as long as contributions from $\Ztwo$ fixed planes and five-branes
are not included. 

\subsection{Orbifold and five-brane contributions}\label{secsixform2}

If one assumes that the Bianchi identity is not $d\G=0$, but instead 
$d\G=\Dfive$ with an exact five-form $\Dfive=d\Dfour$ not depending on
$C_3$, $G_4$ or $C_6$, the action (\ref{S11dsd'}) can be 
consistently modified to become:
\beq
\label{SDfive}
\begin{array}{rcl}
2\kappa_{11}^2\, S&=& 
-{1\over 2}\int_{\Mele}{\G\wedge *\G}
-{1\over 2}\int_{\Mele}{\C\wedge d\C\wedge (\G-\Dfour-{2\over 3}d\C)} 
\crbig
&& +\int_{\Mele}{\Csix\wedge (d\G-\Dfive)}
+ \hbox{Einstein term} +\cdots.
\end{array}
\eeq
The independent fields are $\C$, $\G$ and $\Csix$. The term with
$\Csix$ is modified to obtain the new Bianchi identity with source 
$\Dfive$. The additional $\C\wedge d\C\wedge\Dfour$ term 
is a possible addition to cancel the variation under
(\ref{Mgaugetransf}) of the source contribution $-\Csix\wedge\Dfive$.
We will however see below that if a five-brane is at the origin 
of the source $\Dfive$, another modification arises. 
The dots in action (\ref{SDfive}) denote possible contributions 
which do not involve the 
eleven-dimensional bulk fields and are
related to the dynamics of the magnetic source $\Dfive$.
An example would be the ten-dimensional kinetic terms for the 
gauge fields in a compactification of M-theory on $S^1/\Ztwo$.
The equations of motion for $\C$, $\G$ and $\Csix$ are
\beq
\begin{array}{rclrcl}
\C&:& \hfill d\C\wedge (d\C-\G+\Dfour) &=&
-{1\over 2}\C\wedge (d\G-\Dfive), \hfill\crbig
\G&:& \hfill *\G &=& -d\Csix-{1\over 2}\C\wedge d\C, \hfill\crbig
\Csix&:& \hfill d\G &=& \Dfive. \hfill\crbig
\end{array}
\eeq
Compactification of M-theory on $O_7$ or $S^1/\Ztwo$ has two kinds of 
defects generating sources $\Dfive$: M-theory five-branes and $\Ztwo$ fixed 
planes. A tensor $N=2$ multiplet of massless excitations lives on the 
world-volume of a five-brane \cite{GT, KM} and a $E_8$ super-Yang-Mills 
multiplet is located on each fixed plane \cite{WITTENSTRONG, HW1, HW2}. 

A five-brane in an eleven-dimensional background can be described by
the following world-volume bosonic action \cite{TABRO}
\beq\label{S5b}
S_{\rm M5} = 
\int_{\Wsix}{\L_{\rm kin.}}
-T_5\int_{\Wsix}{\hat\Csix}
-{T_5\over 2}\int_{\Wsix}{\hat\C\wedge d\Btwo} \,,
\eeq
where the hatted fields $\hat\C$ and $\hat\Csix$ are the
eleven-dimensional background fields pulled back onto the
six-dimensional world-volume $\Wsix$ and the two-form $\Btwo$ 
belongs to the $D=6$ supermultiplet of the five-brane\footnote{
See Section \ref{sec5brane} for more detail on this 
multiplet.}.
The first contribution $\L_{\rm kin.}$ describes the kinematics of the 
bosonic degrees of freedom. It includes a Born-Infeld term for the induced
metric tensor $\hat g_{\hat m\hat n}$ coupled to the three-form 
$\Hthree\equiv d\Btwo-\hat\C$, which is submitted to a self-duality 
condition. In the covariant formalism of Pasti, Sorokin and Tonin
\cite{PSTall}, this self-duality condition is generated by  
an auxiliary scalar field. Hence,
${\cal L}_{\rm kin.}$ is a functional of $\hat g_{\hat m\hat n}$, $\Hthree$
and of the auxiliary PST scalar, but its precise form is
unimportant for a while. Notice that invariance under 
(\ref{Mgaugetransf}) of $\Hthree$ implies $\delta\Btwo= \hat\Lambda_2$,
and that, with this transformation of $\Btwo$, the complete action
$S_{\rm M5}$ is gauge invariant.

The five-brane action $S_{\rm M5}$ includes the $\Csix$ term
\beq
\label{delta5def}
-T_5\int_\Wsix\hat\Csix = -T_5\int_\Mele \Csix\wedge\delta_5\,,
\eeq
where the equality for an arbitrary six-form would define 
the closed delta function five-form $\delta_5$.
Comparison with the $\Csix$ term in action (\ref{SDfive}) indicates
that adding a five-brane contribution modifies the source $\Dfive$
according to 
$\Dfive \rightarrow \Dfive + 2\kappa_{11}^2T_5\delta_5$,
without however affecting $\Dfour$ since gauge invariance is obtained
with the new contribution $-{T_5\over2}\int_\Mele \C\wedge d\Btwo\wedge
\delta_5$.

Using $\delta_5$ to rewrite the action (\ref{S5b}) as
\beq\label{S5b2}
S_{\rm M5} = 
\int_{\Mele}{\L_{\rm kin.}\wedge\delta_5}
-T_5\int_{\Mele}{\Csix\wedge\delta_5}
-{T_5\over 2}\int_{\Mele}{\C\wedge d\Btwo\wedge\delta_5}\,,
\eeq
we obtain a complete action from which the modified Bianchi identity with
a five-brane source added can be
deduced as an equation of motion for the six-form $\Csix$:
\beq 
\label{SD5modif}
\begin{array}{rcl}
2\kappa_{11}^2 S&=& 
-{1\over 2}\int_{\Mele}{G_4\wedge *G_4}
-{1\over 2}\int_{\Mele}{C_3\wedge dC_3\wedge (G_4-\Dfour-{2\over 3}dC_3)} 
\crbig
&&
+\int_{\Mele}{C_6\wedge (dG_4-\Dfive- 2\kappa_{11}^2 T_5\delta_5)} \crbig
&&
+ 2\kappa_{11}^2 \int_{\Mele}{\L_{\rm kin.}\wedge\delta_5} 
-\kappa_{11}^2T_5\int_{\Mele}{C_3\wedge d\Btwo\wedge\delta_5}
+\hbox{Einstein term}+\cdots,
\end{array}
\eeq
where the independent fields are the metric, 
$\C$, $\G$, $\Csix$ and $\Btwo$ \footnote{As well as the translational 
degrees of freedom of the five-brane world-volume, in the pull-back
of $\Mele$ onto $\Wsix$.}. Their equations of motion are
\beq
\label{SD5eom}
\begin{array}{rclrcl}
\C&:& \hfill d\C\wedge (d\C-\G+\Dfour) &=&
- 2\kappa_{11}^2 
({\delta\L_{\rm kin.}\over\delta\C}-{T_5\over 2}d\Btwo)\wedge\delta_5
-{1\over 2}\C\wedge (d\G-\Dfive)\,, \hfill\crbig
\G&:& \hfill *\G &=& -d\Csix-{1\over 2}\C\wedge d\C \,, \hfill\crbig
\Csix&:& \hfill d\G &=& \Dfive + 2\kappa_{11}^2 T_5\delta_5 \,, \hfill\crbig
\Btwo&:& \hfill  {\delta\L_{\rm kin.}\over\delta\Btwo}\wedge\delta_5 &=& 
{T_5\over 2}d\C\wedge\delta_5 \,. \hfill
\end{array}
\eeq
Taking the exterior derivative of the first equation and using the
equality
$d({\delta\L_{\rm kin.}\over\delta\C})=
{{\delta\L_{\rm kin.}\over\delta\Btwo}}$
which follows from the fact that $\C$ and $\Btwo$ only appear through
$\Hthree$ in $\L_{\rm kin.}$, we recover the last equation when the third
relation is taken into account.

This discussion can be easily extended to the presence of 
several five-branes. In the case of M-theory on $O_7$, 
five-brane world-volumes must
be embedded in $\Mele$ in a $\Ztwo$-invariant way. 

We now proceed to add the contributions due to $\Ztwo$ fixed planes. 
They will correspond to specific expressions for the source $\Dfive$
and its primitive $\Dfour$ in action (\ref{SD5modif}). And 
since these expressions do not depend on the eleven-dimensional or 
five-brane fields, the equations (\ref{SD5eom}) and their significance 
will remain unchanged. 

In the presence of five-branes, M-theory on $S^1/\Ztwo\times
\Mten$ can be defined by the following Bianchi identity
\cite{WITTENSTRONG, HW1, HW2, WJGP22}:
\beq
\label{MBi}
d\G = -(4\pi)^2{\kappa_{11}^2\over\lambda^2}\biggl[
I_{4,1}\wedge\delta_{1,1} +I_{4,2}\wedge\delta_{1,2}+ \sum_j 
\delta_5({\cal W}_{6,j})\biggr],
\eeq
where ${\cal W}_{6,j}$ is the world-volume of the $j$'th five-brane and
$\delta_5({\cal W}_{6,j})$ the corresponding five-form as defined in
eq. (\ref{delta5def}). The $S^1/\Ztwo$
direction $x^4$ has periodicity $2\pi$, $\Ztwo$ acts according 
to $x^4\rightarrow -x^4$ and fixed points are at $x^4=0$ and $\pi$.
For each five brane with world-volume 
${\cal W}_{6,j}$, there exists a five-brane 
with world-volume given by the image under $\Ztwo$ of ${\cal W}_{6,j}$.
Eq. (\ref{MBi}) also gives the expression $T_5= -8\pi^2/\lambda^2$ 
in terms of the gauge coupling constant $\lambda$ on the ten-dimensional 
fixed planes. The Dirac one-forms on $S^1$ read
\beq
\label{deltas}
\delta_{1,1} = \delta(x^4)\, dx^4, \qquad\qquad
\delta_{1,2} = \delta(x^4-\pi)\, dx^4.
\eeq
Finally, on the ten-dimensional $\Ztwo$ fixed planes, at $x^4=0$ 
and $x^4=\pi$, live four-forms 
\beq
I_{4,i} = {1\over(4\pi)^2}\biggl[{\rm tr}\, F_i^2 - {1\over2}{\rm 
tr}\, R^2\biggr],
\qquad\qquad i=1,2,
\eeq
where each $F_i$ is an $E_8$ gauge curvature and $R$ is Riemann curvature. 
We then conclude that the appropriate bosonic action for M-theory on
$S^1/\Ztwo$ can be written as:
\beq
\label{Sfinal}
\begin{array}{rcl}
S&=& \int_\Mele \, {\cal L}, \crbig
2\kappa_{11}^2\, {\cal L} &=& -{1\over 2}\,G_4\wedge *G_4
-{1\over 2}\, C_3\wedge dC_3\wedge (G_4-\Dfour-{2\over 3}dC_3)
\crbig
&&
+ C_6\wedge \left(dG_4 + (4\pi)^2{\kappa_{11}^2\over\lambda^2}\left[
I_{4,1}\wedge\delta_{1,1} +I_{4,2}\wedge\delta_{1,2}+ \sum_j 
\delta_5({\cal W}_{6,j})\right] \right) \crbig
&&
+2\kappa_{11}^2\sum_j\left( \L_{\rm kin.}(H_{3(j)})
+{4\pi^2\over \lambda^2}\,C_3\wedge d\Btwo_{(j)}\right)
\wedge\delta_5({\cal W}_{6,j}) \crbig
&&
-{\kappa_{11}^2\over\lambda^2}\Bigl( F_1\wedge*F_1\wedge \delta_{1,1} 
+  F_2\wedge*F_2\wedge \delta_{1,2}\Bigr)+\hbox{Einstein term} \,.
\end{array}
\eeq
The last line includes the kinetic terms of the $E_8$ gauge fields
living on each fixed plane and the four-form $\Dfour$ is defined as
the solution to the Bianchi identity (\ref{MBi}) {\it without any 
five-brane}\/:
\beq
\label{Delta4is}
d\Dfour = -(4\pi)^2{\kappa_{11}^2\over\lambda^2}\biggl[
I_{4,1}\wedge\delta_{1,1} +I_{4,2}\wedge\delta_{1,2} \biggr].
\eeq
Notice that each five-brane has its own tensor $H_{3(j)} =
d\Btwo_{(j)}-C_3$, up to the identification of a five-brane 
with its image under $\Ztwo$.

It should be remarked that the theory (\ref{Sfinal}) is not equivalent
to the Ho\v rava-Witten action. It is a generalization of the bosonic 
sector of eleven-dimensional supergravity and it does not include an 
anomaly-cancelling term similar to the contribution 
$\int_{{\cal M}_{11}} C_3 \wedge G_4\wedge G_4$. Cancellation
of chiral anomalies requires the addition of appropriate 
Green-Schwarz counterterms.

\subsection{The background}\label{secback}

The Bianchi identities of M-theory compactified on $O_7$ are 
the components of eq. (\ref{MBi}) reduced on the Calabi-Yau space. 
They are also the field equations of the components of $\Csix$ reduced
on $O_7$. Denoting by $V_6$ the Calabi-Yau volume and
using $\kappa_{11}^2 =  2\pi V_6 \kappa^2$, where $\kappa$ is the 
four-dimensional gravitational constant, one infers that the
dimensionless number $\lambda^2 /V_6$ can be absorbed in the metric 
moduli, so that these identities as well as the four-dimensional 
reduced action depend on a single parameter,
the four-dimensional gravitational constant\footnote{See ref. \cite{DS} 
for a detailed discussion. For the same reason, we may choose 
the $S^1$ radius to be one.}. As mentioned earlier, the field equation
of the component $C_{\mu\nu\rho\sigma i\ov j}$ is the background
equation
\beq
\label{vacuum}
(d\G)_{4ij\ov k\ov l} = -(4\pi)^2{\kappa_{11}^2\over\lambda^2}
\biggl[I_{4,1}\wedge\delta_{1,1} +I_{4,2}\wedge\delta_{1,2}+ \sum_j 
\delta_5({\cal W}_{6,j})\biggr]_{4ij\ov k\ov l}\,.
\eeq
This equation integrated over a closed five-cycle gives, for a globally 
well-defined $G_4$, the standard `cohomology condition' which defines
the embedding of the four-dimensional gauge group into $E_8\times E_8$
\cite{WITTENSTRONG}.  In general, it implies  
non-zero background values for $({\rm tr}\, F_1^2)_{ij\ov k\ov l}$ and/or 
$({\rm tr}\, F_2^2)_{ij\ov k\ov l}$, and relates these vacuum values to 
the Calabi-Yau background $({\rm tr}\, R^2)_{ij\ov k\ov l}$.
Since there are no massless fluctuations associated with this component
of $d\G$, we may assume that the fluctuation $C_{\mu\nu\rho\sigma i\ov j}$ 
is zero when computing the reduced effective Lagrangian, provided we 
develop the theory around the appropriate background. 

We denote the form degree on ${\cal M}_4\times O_7$ as $(m,n,p,q)$. The  
degree on ${\cal M}_4$ is $m$, the holomorphic and anti-holomorphic degrees 
on the Calabi-Yau space are $n$ and $p$, and $q$ is the degree on $S^1/\Ztwo$.
The $SU(3)$ holonomy condition implies that the background $\langle G_4\rangle$
is a $(0,2,2,0)$ form.
The defining equation for the six-form field $C_6$ is the duality equation 
$*G_4 = -dC_6 - {1\over2} C_3\wedge dC_3$. 
The background $\langle dC_6\rangle$ is then a $(4,1,1,1)$
form. By $SU(3)$ holonomy and $\Ztwo$ symmetry,
the background component $\langle C_6\rangle$ of $C_6$ is a $(4,1,1,0)$ form
and $\langle dC_6\rangle = {\partial\over\partial x^4}\langle C_6\rangle
\, dx^4$. The equations defining the background are then\footnote{
When reduced on $\Mfour\times O_7$, $\langle C_3\wedge dC_3\rangle$ vanishes.} 
$\langle dC_6\rangle =-*\langle G_4\rangle$ and
$$
\langle d*d C_6\rangle = 
-(4\pi)^2{\kappa_{11}^2\over\lambda^2}
\Bigl< I_{4,1}\wedge\delta_{1,1} +I_{4,2}\wedge\delta_{1,2}+ \sum_j 
\delta_5({\cal W}_{6,j})\Bigr> \,.
$$
They depend in general of the metric tensor reduced on ${\cal M}_4\times O_7$
since they use the Hodge dual and Dirac tensorial distributions.
This condition has been studied in detail in refs. \cite{LOW1, LOW}.

In our approach based on Lagrangian (\ref{Sfinal}), however, the 
background contribution is more involved. Since the six-form field 
multiplies the Bianchi identity, all $C_6$ background contributions 
automatically cancel. But the background values of $G_4\wedge*G_4$, 
of the Einstein term and of the gauge and brane 
(Born-Infeld) kinetic terms are non-zero. 
We will return to this point when computing the effective four-dimensional
scalar potential, which vanishes, in the next section. 

Our task now is to obtain the four-dimensional Calabi-Yau 
reduction of the action (\ref{Sfinal}), and to extend it to a 
Poincar\'e $N=1$ supergravity. Without five-branes, the result is
well-known either from heterotic strings on $X_6$ \cite{SUGRAhet}
or from M-theory on $O_7$ \cite{NOY, LOW1}, and reference 
\cite{DS} gives a discussion based
on Bianchi identities which is also the approach followed here. 

\section{The M-theory five-brane}\label{sec5brane}
\setcounter{equation}{0}

The action (\ref{Sfinal}) includes kinetic terms for the five-brane 
bosonic degrees of freedom, which in particular propagate 
the self-dual three-form $H_3$. We find useful to incorporate in our
discussion the largest possible symmetry. As a consequence, we
will use for these kinetic terms the formalism of Pasti, Sorokin 
and Tonin \cite{PSTall} adapted to the five-brane \cite{5bPST, 5bCKVP}, 
in which self-duality follows from field equations. 

Since the five-brane has also scalar fields related to the 
translational modes of its world-volume, we begin with a 
brief discussion of the embedding of a world-volume ${\cal W}_6$ in 
$\Mfour\times O_7$. 

\subsection{Reduction of the M-five-brane bosonic action to four-dimensions}

In order to preserve $N=1$ four-dimensional supersymmetry, 
each five-brane must be aligned with a world-volume ${\cal W}_{6,j} = 
\Mfour\times {\cal C}_{2,j}$, the two-cycle ${\cal C}_{2,j}$ in the 
Calabi-Yau manifold being holomorphic \cite{BeckerBS, BVS, WITTENSTRONG}. 
We can then choose the five-brane world-volume coordinates as
\beq\label{yare}
y^{\hat m} = (y^\mu, y, \ov y), \qquad
\hat m=\hat0,\hat1,\ldots,\hat5, \qquad
\mu = 0,1,2,3,
\eeq
with a complex coordinate $y$ along the Calabi-Yau two-cycle.
The embedding of the world-volume in $\Mele$ is defined by
the functions\footnote{Notice that $x^4$ is the 
$S^1/\Ztwo$ direction in $\Mele$ while $y^{\hat4}$ 
(and $y^{\hat5}$) are in the Calabi-Yau manifold.} 
$x^M(y^{\hat m})$, $M=0,1,\ldots,10$, and by the pull-back
functions $ {\partial x^M\over\partial y^{\hat m}}$. In $\Mfour\times O_7$,
we use coordinates $x^M=(x^\mu,x^4,z^i,\ov z^{\ov i})$, $i=1,2,3$, with
$$
x^\mu=y^\mu, 
\qquad z^i=z^i(y^\mu,y), 
\qquad \ov z^{\ov i}=\ov z^{\ov i}(y^\mu,\ov y),
$$
choosing a parametrization of $\Mfour$.

The five-brane excitations are described by a $D=6$ tensor supermultiplet
\cite{GT, KM}. The fields are 
a chiral antisymmetric tensor $\B_{\hat m\hat n}$
(with a self-dual field strength $H_{\hat m\hat n \hat p}$), 
five scalar fields $X_{(1)}, \dots, X_{(5)}$ specifying 
the position of the world-volume $\Wsix$ in $\Mele$, and their fermionic
partners. In our $\Mfour\times O_7$ reduction, we 
neglect the detailed structure of the Calabi-Yau
manifold. Of the five scalar fields, only one survives as the 
massless mode of the Calabi-Yau expansion of $x^4(y^\mu,y,\ov y)$,
\beq
\label{5bembed}
\begin{array}{l}
x^4(y^\mu,y,\ov y) = X(x^\mu) + {\rm massive\,\, modes}, \crbig
z^i = z^i(y),\qquad \ov z^{\ov i} = \ov z^{\ov i}(\ov y), \qquad
x^\mu=y^\mu.
\end{array}
\eeq
This means that we will only retain the following bosonic five-brane
excitations:
\beq\label{5bfields}
\B_{\mu\nu}(x^\mu) = \B_{\mu\nu}(y^\mu), 
\qquad \B_{\hat4\hat5} (x^\mu) =
\B_{\hat4\hat5}(y^\mu) , \qquad
X(x^\mu),
\eeq
and the self-duality condition on $H_{\hat m\hat n\hat p}$ relates 
$\B_{\mu\nu}$ and $\B_{\hat4\hat5}$.
The background value of the scalar field $X$ is 
the five-brane position along the $S^1/\Ztwo$ orbifold direction $x^4$.
Each five-brane generates then in $\Mfour$ two bosonic degrees of 
freedom. By $N=1$ supersymmetry, they will be described by either a 
linear or a chiral multiplet. 

With these choices of embedding and truncation, 
the world-volume induced metric\footnote{
The two-index tensor $g_{MN}$ is the eleven-dimensional metric 
which was defined and used in ref. \cite{DS}. Its reduction can 
also be found in the appendix [eq. (\ref{elemetric})].}
$$
\hat g_{\hat m\hat n} = {\partial x^M\over\partial y^{\hat m}}
{\partial x^N\over\partial y^{\hat n}}g_{MN}
$$
reduces in four dimensions to
\beq\label{inducedg}
\begin{array}{rcl}
\hat g_{\mu\nu} &=& e^{-\gamma-2\sigma}g_{\mu\nu}
+e^{2\gamma-2\sigma}(\partial_\mu X)(\partial_\nu X) , \crbig
\hat g_{\hat4\hat5} &=& k^2e^\sigma, \crbig
\hat g_{\mu\hat4} &=& \hat g_{\mu\hat5} \,\,=\,\, 
\hat g_{\hat4\hat4} \,\,=\,\, \hat g_{\hat5\hat5} \,\,=\,\, 0, 
\end{array}
\eeq
where 
$k^2 =\delta_{i\ov i}{\partial z^i\over\partial y}
{\partial\ov z^{\ov i}\over\partial\ov y}$ is a constant 
(a background value) in our Kaluza-Klein truncation. 

To describe the dynamics of the bosonic fields (\ref{5bfields}) 
and their couplings to four-dimensional supergravity,
we need an action for the five-brane coupled to eleven-dimensional 
supergravity. 
Using the PST formalism to write covariant Lagrangians for self-dual 
(or anti-self-dual) tensors, 
a kappa-symmetric covariant world-volume Lagrangian for the 
five-brane excitations has been constructed \cite{5bPST, 5bCKVP}, 
completing earlier work \cite{earlier, WJGP22, TABRO}. 
In a non-trivial 
eleven-dimensional supergravity background, the action has
two parts: a kinetic Lagrangian with a Born-Infeld term 
involving the three-index tensor $H_{\hat m\hat n\hat p}$ 
and a Wess-Zumino
term involving both $\C$ and its dual $\Csix$. The bosonic action 
is\footnote{Our conventions are mostly as 
in ref. \cite{5bCKVP}. We consider here a single five-brane.}:
\beq\label{brane1}
\begin{array}{rcl}
\S_{\rm M5} &=& T_5
\bigint_{\Wsix} d^6y \left( -\sqrt{-\det(\hat g_{\hat m\hat n}
+ i{H^*}_{\hat m\hat n})}
-{1\over4}\sqrt{-\hat g}\,{\cal V}_{\hat l} 
H^{*\hat l\hat m\hat n}H_{\hat m\hat n\hat p}{\cal V}^{\hat p}\right) \crbig
&&-T_5\bigint_{\Wsix} \left( \hat\Csix -{1\over2}d\Btwo\wedge 
\hat\C \right).
\end{array}
\eeq
The second line is as in eq. (\ref{S5b}) and the first two terms define 
the kinetic Lagrangian in the PST formalism.
In this expression, $\hat\C$ and $\hat\Csix$ are the eleven-dimensional
background fields pulled back onto the world-volume using derivatives
$\partial x^M/\partial y^{\hat m}$,
\beq\label{Cpulled}
\hat C_{\hat m\hat n\hat p} = {\partial x^M\over\partial y^{\hat m}}
{\partial x^N\over\partial y^{\hat n}}{\partial x^P\over\partial 
y^{\hat p}}C_{MNP}, 
\qquad\qquad
\hat C_{\hat m_1\ldots \hat m_6} = {\partial x^{M_1}\over\partial 
y^{\hat m_1}}\ldots
{\partial x^{M_6}\over\partial y^{\hat m_6}}C_{M_1\ldots M_6},
\eeq
$T_5$ is the brane tension and
\beq\label{branedefs}
\begin{array}{rcl}
H_{\hat m\hat n\hat p} &=& 3\,\partial_{[\hat m}\B_{\hat n\hat p]} 
- \hat C_{\hat m\hat n\hat p}, \crbig
{H^*}_{\hat m\hat n} &=& H^*_{\hat m\hat n\hat p}{\cal V}^{\hat p} , \crbig
H^{*\hat m\hat n\hat p} &=& -{1\over3!\sqrt{-\hat g}}\,\epsilon^
{\hat m\hat n\hat p\hat q\hat r\hat s}H_{\hat q\hat r\hat s}, \crbig
d\Btwo &=& {1\over2}\,\partial_{\hat m} \B_{\hat n\hat p}
\, dy^{\hat m}\wedge dy^{\hat n}\wedge dy^{\hat p}.
\end{array}
\eeq
Finally, 
\beq\label{Vis}
{\cal V}_{\hat m} ={\partial_{\hat m} A\over\sqrt{(\partial_{\hat n} 
A)(\partial^{\hat n}A)}}, \qquad
({\cal V}_{\hat m}{\cal V}^{\hat m} = 1),
\eeq
where $A(y^{\hat m})$ is the auxiliary scalar field introduced by PST to 
impose the self-duality of the tensor 
$H_{\hat m\hat n\hat p}$ as an equation of motion. 

Since we will consider only four-dimensional contributions with up to
two derivatives, it will be sufficient to write
\beq\label{det2der}
\sqrt{-\det(\hat g_{\hat m\hat n}+i{H^*}_{\hat m\hat n})} \simeq
\sqrt{-\hat g}\left( 1-{1\over4} H^{*\hat m\hat n}{H^*}_{\hat m\hat n} \right).
\eeq
The action (\ref{brane1}) simplifies then to
\beq\label{brane2}
\begin{array}{rcl}
\S_{\rm M5} &=& -T_5
\bigint_{\Wsix} d^6y \sqrt{-\hat g}\left( 
{1\over4}{\cal V}_{\hat l}H^{*\hat l\hat m\hat n}(H_{\hat m\hat n\hat p}
-{H^*}_{\hat m\hat n\hat p}){\cal V}^{\hat p} + 1\right) \crbig
&& -T_5\bigint_{\Wsix} \left( \hat\Csix -{1\over2}d\Btwo\wedge 
\hat\C \right).
\end{array}
\eeq
The PST formalism possesses various local symmetries. 
One of them allows a gauge choice in which $A(y^{\hat m})$ 
is a function of $y$ and $\ov y$ only, so that 
\beq\label{Vgauge}
{\cal V}^\mu = 0, \qquad
{\cal V}_{\hat4}{\cal V}^{\hat4} + {\cal V}_{\hat5}{\cal V}^{\hat5} = 1,
\eeq
which preserves four-dimensional Lorentz covariance. With our truncation 
(\ref{5bfields}) of the five-brane excitations and of the bulk
fields, we are led to only retain components
\beq\label{Hcomp}
\begin{array}{c}
H_{\mu\hat4\hat5}= \partial_\mu\B_{\hat4\hat5}
-\hat C_{\mu\hat4\hat5}, \qquad 
H_{\mu\nu\rho} = 3\partial_{[\mu}\B_{\nu\rho]}- \hat C_{\mu\nu\rho}, 
\crbig
\B_{\hat4\hat5}\equiv k^2\B,\qquad
\hat C_{\mu\hat4\hat5} = k^2a(x)\partial_\mu X, \qquad
\hat C_{\mu\nu\rho} =  3C_{[\mu\nu4}\,\partial_{\rho]}X,
\end{array}
\eeq
where $a(x)$ is defined by $C_{i\ov j4}=ia(x)\delta_{i\ov j}$.
In addition, our reduction of the eleven-dimensional space-time metric
(\ref{elemetric}) implies that
\beq\label{hatgis}
\sqrt{-\hat g} \simeq k^2 e \, e^{-2\gamma-3\sigma}\,
\left( 1+{1\over2}e^{3\gamma}(\partial_\mu X)(\partial^\mu X) \right),
\eeq
where $e^2=-\det(g_{\mu\nu})$ is now the determinant 
of the four-dimensional space-time metric.

The reduction of the term involving the six-form field follows from
two facts. Firstly, with the embedding (\ref{5bembed}) of ${\cal W}_6$ 
into $\Mele$, one can write
$$
\hat C_{\mu\nu\rho\sigma \hat4\hat5} = -i{\partial z^i\over\partial y}
{\partial\ov z^{\ov j}\over\partial\ov y}
\left[ \langle C\rangle_{\mu\nu\rho\sigma i\ov j} +
C_{\mu\nu\rho\sigma i\ov j} +
4(\partial_{[\mu} X)C_{4\nu\rho\sigma] i\ov j}\right],
$$
where $\langle C\rangle_{\mu\nu\rho\sigma i\ov j}$ is the background 
contribution discussed in paragraph \ref{secback} and 
$C_{\mu\nu\rho\sigma i\ov j}$ is the field fluctuation.
Notice that the equations defining this background involve the reduced 
eleven-dimensional metric\footnote{In particular in the Hodge dual.}
and $\langle C\rangle_{\mu\nu\rho\sigma i\ov j}$ does depend on the 
metric moduli $\sigma$ and $\gamma$. 
Secondly, since $\Csix$ is even under $\Ztwo$, $C_{4\nu\rho\sigma i\ov j}$ 
is cancelled by the $O_7$ reduction and the component $C_{\mu\nu\rho\sigma
i\ov j}$ generates the background equation and can be omitted in the 
four-dimensional effective Lagrangian. 

The four-dimensional five-brane action reads then
\beq\label{branefinal}
\begin{array}{rcl}
\S_{\rm M5} &=& \bigint_{\Mfour}d^4x\, \L_{\rm M5}, \crbig
\L_{\rm M5} &=& -{\tilde T\over2} \biggl[
{1\over3!} e \,e^{\gamma+3\sigma}H_{\mu\nu\rho}H^{\mu\nu\rho}
-{1\over3!}\epsilon^{\mu\nu\rho\sigma}
\Bigl( \partial_\mu\B-(\partial_\mu X)\,a\Bigr)H_{\nu\rho\sigma} 
\crbig
&& -{1\over2} \epsilon^{\mu\nu\rho\sigma}(\partial_\mu\B)
(\partial_\nu X) C_{\rho\sigma4} -{1\over2} \epsilon^{\mu\nu\rho\sigma}
\,a\,(\partial_\mu \B_{\nu\rho})(\partial_\sigma X) \crbig
&& 
+ e \,e^{\gamma-3\sigma}(\partial_\mu X)(\partial^\mu X)
+ 2e \,(e^{-2\gamma-3\sigma}+\langle C\rangle)\biggr].
\end{array}
\eeq
This derivation uses 
$$
T_5\,\int_{\Wsix}d^6y\,\sqrt{-\hat g}\,(\ldots) = 
\tilde T\,\int_{\Mfour}d^4x\, e \, e^{-2\gamma-3\sigma}\,
\left( 1+{1\over2}e^{3\gamma}(\partial_\mu X)(\partial^\mu X) \right)(\ldots),
$$
where
$$
{\tilde T\over T_5} = \int_{\Ctwo} dy\,d\ov y\,\, {\partial z^i\over\partial y}
{\partial\ov z^{\ov j}\over\partial\ov y} \delta_{i\ov j}
$$
is the volume of the holomorphic two-cycle in the Calabi-Yau manifold,
and the definition $\langle C\rangle_{\mu\nu\rho\sigma i\ov j} =
i e \epsilon_{\mu\nu\rho\sigma} \langle C\rangle \delta_{i\ov j}$.
The scalar field $\B$ acts as a Lagrange multiplier. It 
imposes the constraint
$$
\epsilon^{\mu\nu\rho\sigma}\partial_\mu\Bigl( 
H_{\nu\rho\sigma}+3(\partial_\nu X)C_{\rho\sigma4} \Bigr) =
\epsilon^{\mu\nu\rho\sigma}\partial_\mu\left( 
H_{\nu\rho\sigma} + \hat C_{\nu\rho\sigma} \right) = 0.
$$
Its solution is the second eq. (\ref{Hcomp}). We can then consider 
the Lagrangian (\ref{branefinal}) as a function of the unconstrained fields
$H_{\mu\nu\rho}$, $X$ and $\hat\B=\B-Xa$:
\beq\label{branefinal2}
\begin{array}{rcl}
\L_{\rm M5} &=& -{\tilde T\over2} \biggl[
{1\over3!} e \,e^{\gamma+3\sigma}H_{\mu\nu\rho}H^{\mu\nu\rho} 
+ e \,e^{\gamma-3\sigma}(\partial_\mu X)(\partial^\mu X)
\crbig
&&+{1\over3}X\epsilon^{\mu\nu\rho\sigma} H_{\mu\nu\rho}(\partial_\sigma
a) + {1\over2}X^2 \epsilon^{\mu\nu\rho\sigma}(\partial_\mu a)
(\partial_\nu C_{\rho\sigma4})
\crbig
&&
-{1\over3!}\epsilon^{\mu\nu\rho\sigma}(\partial_\mu\hat\B)
\Bigl(H_{\nu\rho\sigma} - 3X(\partial_\nu C_{\rho\sigma4})\Bigr)
+2e \,(e^{-2\gamma-3\sigma}+\langle C\rangle)\biggr].
\end{array}
\eeq
The last term seems to indicate the presence of a scalar potential. 
However, solving the equation defining the six-form background field
shows a cancellation: the scalar potential vanishes as expected by 
the stability of the configuration which is protected by the residual 
supersymmetry \cite{TDS}. We will see in paragraph \ref{secsusy2} that 
the supermultiplet structure required to supersymmetrize this bosonic 
action does not allow the presence of a scalar potential.

The Lagrangians (\ref{branefinal}) and (\ref{branefinal2})
are invariant under the residual symmetries: 
\beq\label{sym1}
\begin{array}{c}
\delta C_{\mu\nu4} = 2\partial_{[\mu}\Lambda_{\nu]}, \crbig
\delta a = c, \qquad \delta\B=cX, \qquad
\delta\hat\B=0, \qquad
c = {\rm constant}.
\end{array}
\eeq
Note moreover that $\B$ appears in the Lagrangian (\ref{branefinal}) 
only through its derivatives, so the independent symmetry
\beq
\delta\B=c', \qquad c' = {\rm constant},
\eeq
is also present.
Solving for $\hat\B$ in eq. (\ref{branefinal2})
leads to a Lagrangian for $\B_{\mu\nu}$ 
and $X$, which will be supersymmetrized using a linear multiplet. 
And solving for $H_{\mu\nu\rho}$ leads to a theory containing a chiral
multiplet with scalar components $X$ and $\hat\B$. This chiral--linear
duality is the four-dimensional consequence of the self-duality of the
brane three-index tensor $H_{\hat m\hat n\hat p}$, when expressed
in the covariant formalism of PST. 

We now consider the supersymmetrization in four space-time dimensions 
of the reduced five-brane Lagrangian, first without supergravity 
background, then with the coupling to the eleven-dimensional 
background fields.

\subsection{Supersymmetrization without supergravity background}

Our first goal is to identify the supermultiplet content of 
the effective four-dimensional
supergravity expected to arise from our truncation of the five-brane 
spectrum. The simplest procedure is to consider the flat, zero-background
limit of the five-brane Lagrangian (\ref{branefinal2}), which 
becomes
\beq\label{braneflat1}
\L_{\rm M5,\,flat} = -{\tilde T\over2} \left[
{1\over3!} H_{\mu\nu\rho}(H^{\mu\nu\rho}
+\epsilon^{\mu\nu\rho\sigma}\,\partial_\sigma\B) 
+ (\partial_\mu X)(\partial^\mu X)\right].
\eeq
Introducing for convenience the four-dimensional vector field
\beq\label{vh}
v^\mu = {1\over3!}\epsilon^{\mu\nu\rho\sigma}H_{\nu\rho\sigma}\,,
\eeq
we obtain
\beq\label{5bactflat}
\L_{\rm M5,\,flat} = {\tilde T\over2} \Bigl[v^\mu 
\left(\partial_\mu\B+v_\mu\right) 
-(\partial_\mu X)(\partial^\mu X)\Bigr].
\eeq
Solving for $v_\mu$ leads to $v_\mu=-{1\over2}\partial_\mu\B$, so that
\beq\label{5bflat2}
\L_{\rm M5,\,flat} = -{\tilde T\over2} \left[{1\over4}
(\partial_\mu\B)(\partial^\mu\B) 
+ (\partial_\mu X)(\partial^\mu X)\right].
\eeq
Alternatively, solving for $\B$ gives
\beq\label{bmunu}
\partial_\mu v^\mu=0 \qquad\longrightarrow\qquad 
H_{\mu\nu\rho} = 3\partial_{[\mu}\B_{\nu\rho]},
\eeq
and we obtain the equivalent form of the Lagrangian
\beq\label{5dflat3}
\L_{\rm M5,\,flat} = -{\tilde T\over2} \left[
{1\over3!}H^{\mu\nu\rho}H_{\mu\nu\rho}
+(\partial_\mu X)(\partial^\mu X)\right].
\eeq
This discussion illustrates again how the six-dimensional self-duality 
condition on $H_{\hat m\hat n\hat p}$ translates in the 
truncated four-dimensional theory into a duality equivalence of 
an antisymmetric tensor $\B_{\mu\nu}$ with a (pseudo)scalar
$\B$. 

We now observe that expression (\ref{5bactflat}) 
is precisely the bosonic part of the supersymmetric Lagrangian
\beq\label{5bactflat2}
\L_{\rm flat} = -\tilde T
\int d^2\theta d^2\ov\theta\, \left( \hat V^2 -{1\over2}(\hat S+\ov{\hat S})
\hat V \right),
\eeq 
where $\hat V$ is a real vector superfield and $\hat S$ is a chiral 
superfield. Using the component expansions
\beq\label{VhatSexp}
\begin{array}{rcl}
{\hat V} &=& \hat C + (\theta\sigma^\mu\ov\theta) \hat v_\mu 
+\theta\theta (\hat m+i\hat n)+\ov{\theta\theta}(\hat m-i\hat n) \crbig
&& +\,\theta\theta\ov{\theta\theta}(\hat d-{1\over4}\Box \hat C)
+\cdots \, ,
\crbig
\hat S &=& \hat s -\theta\theta\hat f_s 
-i(\theta\sigma^\mu\ov\theta) \partial_\mu\hat s
+{1\over4}\theta\theta\ov{\theta\theta}\Box\hat s +\cdots \, , 
\end{array}
\eeq
where the dots indicate fermion contributions, 
the bosonic part of the supersymmetric Lagrangian 
(\ref{5bactflat2}) is
\beq\label{5bactflat3}
\begin{array}{rcl}
\L_{\rm flat,\,bos.} &=& {\tilde T\over2} \Bigl[
\hat v^\mu(\hat v_\mu-\partial_\mu\Im\hat s)
-(\partial_\mu\hat C)(\partial^\mu\hat C) \crbig
&& -2\hat d(2\hat C-\Re\hat s)
-4(\hat m^2+\hat n^2) - \left(\hat f_s(\hat m-i\hat n) +{\rm
c.c.}\right)\Bigr], 
\end{array}
\eeq
omitting a space-time derivative. The second line is auxiliary and vanishes
when solving for either $\Re\hat s$ and $\hat f_s$ or $\hat m$,
$\hat n$, $\hat d$ and $\hat f_s$. The first line is eq. (\ref{5bactflat}).

The chiral-linear duality present in the globally supersymmetric 
Lagrangian (\ref{5bactflat2}) is the consequence,
in the truncated theory, of the self-duality property 
of the five-brane antisymmetric tensor. Explicitly, 
solving for the vector superfield $\hat V$ in eq. (\ref{5bactflat2}) leads
to $\hat V={1\over4}(\hat S+\ov{\hat S})$. For the bosonic components,
this is $\hat C={1\over2}\Re\hat s$, 
$\hat v_\mu= {1\over2}(\partial_\mu\Im\hat s)$
and $\hat m+i\hat n=-{1\over4}\hat f_s$. The supersymmetric Lagrangian
becomes then
$$
\L_{\rm flat}= 
{\tilde T\over8} \Dint \hat S\ov{\hat S} =
-{\tilde T\over8}\left[(\partial_\mu\hat s)(\partial^\mu\ov{\hat s}) -
\hat f_s\ov{\hat f_s}\right] +{\rm fermionic\,\,terms}. 
$$
Alternatively, we can rewrite expression (\ref{5bactflat2}) as
$$
\L_{\rm flat} = -\tilde T\Dint \hat V^2
-{\tilde T\over8}\Fint\hat S\ov{\cal DD}\hat V
-{\tilde T\over8}\Fbarint \ov{\hat S}{\cal DD}\hat V\,,
$$
and solve for the chiral superfield $\hat S$, implying that $\hat V$
is a real linear multiplet 
$\hat L$, $\ov{\cal DD}\hat L={\cal DD}\hat L=0$. 
For the bosonic components, solving 
for $\hat s$ and $\hat f_s$ in expression (\ref{5bactflat3}) leads
to $\hat d=\hat m=\hat n=0$ and
$$
\partial^\mu \hat v_\mu=0 \qquad\longrightarrow\qquad
\hat v_\mu= -{1\over3!}\epsilon_{\mu\nu\rho\sigma}H^{\nu\rho\sigma}
=-{1\over2}\epsilon_{\mu\nu\rho\sigma}\partial^\nu b^{\rho\sigma}.
$$
The Lagrangian becomes
$$
\L_{\rm flat} = -\tilde T\Dint \hat L^2 = 
-{\tilde T\over2}\Bigl[(\partial_\mu\hat C)(\partial^\mu \hat C)
+{1\over3!}H^{\mu\nu\rho}H_{\mu\nu\rho}\Bigr] +{\rm fermionic\,\,terms.}
$$

\subsection{Supersymmetrization with supergravity background}\label{secsusy2}

We now turn on the supergravity background and return to Lagrangian
(\ref{branefinal2}) to derive its supersymmetric extension. 

The description in terms of superconformal multiplets of the supergravity
bulk fields has been discussed in detail in ref. \cite{DS}. The dilaton
and universal modulus are respectively described by two vector multiplets, 
$V$ with weights $\omega=2$, $n=0$ and $V_T$ with zero weights. Bianchi
identities in $\Mele$ would constrain $V$ to be linear and $V_T$ to
be $T+\ov T$ in terms of a chiral multiplet $T$.
Writing the (bosonic) component expansions as
$$
\begin{array}{rcl}
V &=& (C,0,H,K,v_\mu,0, d-\Box C-{1\over3}CR), 
\crbig 
V_T &=& (C_T,0,H_T,K_T,T_\mu,0, d_T-\Box C_T), \crbig 
\end{array}
$$
the identification is
\beq
\label{identif}
\begin{array}{rclrcl}
4\kappa^2 C &=& {\lambda^2\over V_6} e^{-3\sigma}\,, & \qquad 
4\kappa^2 v_\mu &=& {\lambda^2\over V_6}\,{e\over2}
\epsilon_{\mu\nu\rho\sigma}\partial^\nu C^{\rho\sigma4}\,, 
\crbig
C_T &=& 2{\lambda^2\over V_6}e^\gamma \,, &\qquad 
T_\mu &=& -2{\lambda^2\over V_6}\partial_\mu a\,.
\end{array}
\eeq
Since we may redefine the dimensionless quantity $\lambda^2/V_6$ by a
scaling of the moduli, we take $\lambda^2/V_6=1$ from here on. 
To describe the five-brane degrees of freedom, we introduce as in the 
previous paragraph two supermultiplets: a vector supermultiplet 
$\hat V$ and a chiral supermultiplet $\hat S$. We choose them with 
zero conformal and chiral weights ($\omega=n=0$). Their bosonic 
component expansions are
$$
\begin{array}{rcl}
\hat V &=& (\hat C, 0, \hat H, \hat K, \hat v_\mu, 0, \hat d-\Box\hat C), 
\crbig
\hat S &=& ( \hat s, 0, -\hat f_s, i\hat f_s, i\partial_\mu\hat s,0,0).
\end{array}
$$
To bring the Lagrangian (\ref{branefinal2}) in a form appropriate for 
supersymmetrization in terms of $\hat V$, $\hat S$, $V$ and $V_T$, we
observe that the dimensions of the brane fields $\hat{\cal B}$, $X$ and
$H_{\mu\nu\rho}$ (which are $-1$, $-1$ and $0$) do not fit with 
those of components $\hat s$, $\hat C$ and $\hat v_\mu$ ($0$, $0$ and $1$). 
Since the only scale in our four-dimensional 
Poincar\'e supergravity should be $\kappa$, we first introduce a dimensionless
five-brane coupling constant
\beq
\label{tauis}
\tilde T = {\tau\over\kappa^4},
\eeq
and perform the rescalings
\beq
\label{rescale}
H_{\mu\nu\rho} = \kappa \tilde H_{\mu\nu\rho}\,,\qquad
X = \kappa \tilde X \,,\qquad
\hat{\cal B} = \kappa\tilde{\cal B} \,.
\eeq
Action (\ref{branefinal2}) rewrites then as
\beq
\label{newS}
\begin{array}{rcl}
\L_{\rm M5} &=& -{\tau\over2\kappa^2} \biggl[
{1\over3!} e \,e^{\gamma+3\sigma}\tilde H_{\mu\nu\rho}\tilde H^{\mu\nu\rho} 
+ e \,e^{\gamma-3\sigma}(\partial_\mu\tilde X)(\partial^\mu\tilde X) \crbig
&&+{1\over3}\tilde X\epsilon^{\mu\nu\rho\sigma} 
\tilde H_{\mu\nu\rho}(\partial_\sigma a)
+{1\over2}\tilde X^2 \epsilon^{\mu\nu\rho\sigma}(\partial_\mu a)
(\partial_\nu C_{\rho\sigma4})
\crbig
&&
-{1\over3!}\epsilon^{\mu\nu\rho\sigma}(\partial_\mu\tilde\B)
\Bigl(\tilde H_{\nu\rho\sigma} - 3\tilde X(\partial_\nu C_{\rho\sigma4})\Bigr)
\biggr] - V_0 \,,
\end{array}
\eeq
with an apparent scalar potential
\beq
\label{V0is}
V_0 = {\tau\over\kappa^4} \,e\,(e^{-2\gamma-3\sigma}+\langle C\rangle) \,.
\eeq
Then, to go to the superconformal formalism, we recall \cite{DS} that 
${1\over\kappa^2}$ is the Poincar\'e gauge-fixed value of the 
multiplet\footnote{The chiral $S_0$, with weights $w=n=1$, is the 
compensating multiplet: some of its components are used to gauge fix
superconformal symmetries to realize Poincar\'e invariance only.}
\beq
\label{Upsilonis}
\Upsilon = (S_0\ov S_0 V_T)^{3/2} (2V)^{-1/2}.
\eeq
Suppose that we identify the scalar field $\tilde X$ with the lowest component
$\hat C$ of the brane multiplet $\hat V$. Identifications (\ref{identif})
also indicate that $e^{-3\sigma}$ is the lowest component 
of $4V\Upsilon^{-1}$ while $e^\gamma$ is the lowest component
of ${1\over2}V_T$. We then infer that the first line of action (\ref{newS}) 
appears in the component expansion of 
$$
-\tau[VV_T\hat V^2]_D \,,
$$
which is independent from $\Upsilon$.
Comparison of the $\hat v^\mu\hat v_\mu$ term with the 
$\tilde H_{\mu\nu\rho}\tilde H^{\mu\nu\rho}$ term in the actions 
leads then to the identifications
\beq
\label{hatVfields}
\hat C=\tilde X,\qquad \hat v_\mu = -{e\over6}
{1\over4\kappa^2 C}\epsilon_{\mu\nu\rho\sigma}
\tilde H^{\nu\rho\sigma}\,. 
\eeq
With these results, the vector component of $V\hat V$ is then
$$
C\hat v_\mu + \hat C v_\mu = -{1\over4\kappa^2}
{e\over3!}\epsilon_{\mu\nu\rho\sigma}(\tilde H^{\nu\rho\sigma} 
- 3\tilde X \partial^\nu C^{\rho\sigma4} ) \,,
$$
which is the combination appearing in the last line of Lagrangian
(\ref{newS}). We then conclude that 
\beq\label{Lbraneis}
\L_{\rm brane} = -\tau\left[ VV_T\hat V^2
- {1\over2}(\hat S+\ov{\hat S})V\hat V \right]_D
\eeq
is the superconformal tensor calculus expression for the five-brane
kinetic Lagrangian, with in addition
\beq
\label{hatBis}
\hat{\cal B} = {\rm Im}\, \hat s \,.
\eeq
Expression (\ref{Lbraneis}) is independent from the compensating 
multiplet $S_0$ and completely frame-inde\-pen\-dent.
Its component expansion does not include any $eR$ term and 
the Einstein frame condition for dilatation symmetry would not be 
affected by its addition to bulk (and $S^1/\Ztwo$ plane) contributions.
The bosonic component expression reads
\beq\label{Lbranecomp}
\begin{array}{rcl}
e^{-1}\L_{\rm brane} &=&
-\tau CC_T\left((\partial_\mu\hat C)(\partial^\mu\hat C) 
-\hat v_\mu\hat v^\mu \right) 
+ 2 \tau C\hat C\hat v^\mu T_\mu
+ \tau\hat C^2 v^\mu T_\mu 
\crbig
&& 
+\tau (\partial_\mu\Im \hat s) (C\hat v^\mu+\hat C v^\mu)
\crbig
&& 
+\tau\hat C^2(C_Td - Cd_T) 
-2\tau C\hat C(\partial_\mu C_T)(\partial^\mu\hat C)
-\tau\hat C^2(\partial_\mu C)(\partial^\mu C_T)
\crbig
&& 
+\tau (\Re\hat s - 2\hat CC_T)
\left( C\hat d+\hat Cd-v^\mu\hat v_\mu +
(\partial^\mu C)(\partial_\mu\hat C) \right) 
\crbig
&&
+ e^{-1}\L_{\rm aux.} 
+ \hbox{total derivative}.
\end{array}
\eeq
The auxiliary Lagrangian vanishes `on-shell': it is a quadratic expression in
$H$, $K$, $H_T$, $K_T$, $\hat H$, $\hat K$ and $\hat f_s$. To compare the
above expression with eq. (\ref{newS}), we also need to solve for
$\Re\hat s$, which is not generated by the reduction of the brane bosonic
world-volume action: its presence is required by supersymmetry only. 
The fourth line is then eliminated. All contributions from the third line 
are related by supersymmetry to propagation of the background fields
and are invisible in eq. (\ref{newS}). And the first two lines with 
identifications (\ref{hatVfields}) and (\ref{hatBis}) correspond to
eq. (\ref{newS}), with the exception of the scalar potential $V_0$
which cannot arise from the superconformal expression (\ref{Lbraneis}).

As expected from the self-duality of the brane tensor $H_{\hat m
\hat n\hat p}$, the supergravity Lagrangian (\ref{Lbraneis}) has 
chiral-linear duality. Solving for the vector superfield $\hat V$ gives 
$\hat V={1\over 4}V_T^{-1}(\hat S + \ov{\hat S})$ and we obtain
$$
\L_{\rm brane,\,chiral} = {\tau\over 16} 
\left[ V V_T^{-1} (\hat S + \ov{\hat S})^2 \right]_D.
$$
Alternatively, solving for the scalar superfield $\hat S$ leads to 
$V\hat V=\hat L$, where $\hat L$ is a real linear 
superfield, and then
$$
\L_{\rm brane,\,linear} = -\tau \left[ V_T V^{-1} {\hat L}^2 \right]_D.
$$
Chiral-linear duality requires invariance under $\delta\hat S =$ an imaginary
constant. This symmetry also excludes a superpotential and then
the generation of a scalar potential. 

The conclusion is that the superconformal Lagrangian (\ref{Lbraneis})
provides the four-dimensional effective kinetic Lagrangian for the 
brane modulus multiplet. As in action (\ref{Sfinal}), the complete effective 
four-dimensional supergravity is the known effective theory of orbifold gauge 
and matter multiplets plus expression (\ref{Lbraneis}). Most importantly, the 
brane contributions to the background equations must be taken into account to 
correctly evaluate the scalar potential. This is the last point we need to 
discuss before analysing the complete supergravity theory. 

\subsection{Background and scalar potential}

Returning to the bosonic action (\ref{Sfinal}), we observe that the 
background value of the eleven-dimensional
Lagrangian is
$$
\langle{\cal L}\rangle = \biggl\langle
-{1\over2\kappa_{11}^2}\left[ eR + {1\over2}
G_4\wedge*G_4 \right] + {\cal L}_{\rm kin.}(H_3)\wedge \delta_5({\cal W}_6)
\biggr\rangle \,,
$$
for a single brane and omitting gauge field contributions on orbifold planes. 
In our reduced metric, it is in principle a function of the background
scalar fields $\sigma(x^4)$ and $\gamma(x^4)$, and of their first and 
second derivatives which appear in the curvature scalar $R$. 
However, using the conditions imposed by the background Einstein equations,
one finds that $\langle{\cal L}\rangle$ is a derivative,
$$
\langle{\cal L}\rangle = {1\over2\kappa^2_{11}}\,
{d\over dx^4}\left[ e^{-3\gamma}{d\over dx^4}(\gamma+2\sigma) \right]\,,
$$
which disappears after integration on $x^4$: the 
four-dimensional effective Lagrangian has zero background value. 
As a result, the effective four-dimensional scalar potential generated 
by the brane modulus vanishes. 

Taking several branes and the orbifold planes into account leads to 
the same result: the scalar potential vanishes as long as a superpotential 
is not generated by charged matter chiral superfields. 

\section{The coupled theory}\label{sec5bcoupl}
\setcounter{equation}{0}

In compactified M-theory, the presence of the five-brane modulus multiplet 
does not modify the Bianchi identities verified by the massless components
$G_{4\mu\nu\rho}$, $G_{4\mu i\ov j}$ and $G_{4ijk}$. Its effect on the 
four-dimensional effective supergravity is simply the
addition of the kinetic Lagrangian (\ref{Lbraneis}) and 
the modification of the background equation (\ref{vacuum}) by 
the source terms proportional to 
$\delta_5({\cal W}_{6,j})$. More changes will occur 
with gauge thresholds and anomaly-cancelling terms, which 
can be regarded as `higher-order' corrections.

The complete effective supergravity\footnote{Up to terms with two 
derivatives.} of M-theory compactified on 
$O_7$ with a five-brane can be written as follows:
\beq\label{Lcomplete}
\begin{array}{rcl}
\L &=& 
\left[-(S_0\overline{S_0}V_T)^{3/2}(2V)^{-1/2}
-(S+\overline S)(V+2\Omega) 
+(U(W-\alpha M^3)+{\rm c.c.})\right. \crbig
&& 
+\left.(L_T-2\sum_a{\beta^a\Omega^a})(V_T+2\overline Me^AM)
+V(\epsilon|\alpha M^3|^2-2\delta\overline Me^AM)\right]_D
\crbig
&&
+\left[S_0^3W\right]_F
\crbig
&&-\tau\left[VV_T\hat V^2
-{1\over 2}(\hat S+\overline{\hat S})V\hat V\right]_D
+{\tau\over4}\left[ \hat S\sum_a \hat\beta^a {\cal W}^a{\cal W}^a \right]_F,
\end{array}
\eeq
The first three lines \cite{DS} collect all contributions from gauge
multiplets (in the Chern-Simons superfields $\Omega^a$, $\Omega=\sum_a 
{c^a\Omega^a}$, for a gauge group $\prod_a G^a$) and charged matter 
multiplets\footnote{The chiral multiplet $M$ denotes a generic charged 
matter multiplet, for instance a {\bf 27} of an $E_6$ gauge group.} $M$. 
They also include the contributions of the massless
modes of $G_4$ and of the metric tensor, in the multiplets $V$, $V_T$
and $W$. The first term is the bulk Lagrangian \cite{CFV, DQQ}
produced by the reduction 
of the CJS theory. The next terms induce by the field equations 
of the Lagrange multipliers $S$, $U$ and $L_T$ the Bianchi identities. 
The solutions are: 
$$
\begin{array}{rrcll}
S \,\,({\rm chiral}, w=n=0):&\quad V &=& L-2\Omega \quad  
&(L {\rm\,\, linear}, w=2, n=0),
\crbig
U \,\,({\rm vector}, w=2, n=0):&\quad W &=& \alpha M^3 + ic \quad 
&(c {\rm\,\, real}),
\crbig
L_T \,\,({\rm linear}, w=2, n=0):&\quad 
V_T &=& T+\ov T -2\ov Me^A M \,\,  &(T {\rm\,\, chiral}, w=n=0).
\end{array}
$$
Reduction of the action (\ref{Sfinal}) shows that the massless 
components of the six-form field are included in these Lagrange 
multipliers.
The single term in the third line is the superpotential, as defined by 
the Bianchi identity induced by $U$.
The contributions with coefficients $\beta^a$, $\epsilon$ and $\delta$ 
are higher-order corrections following from anomaly cancellation. They 
generate in particular gauge thresholds. This formulation is derived 
and explained in ref. \cite{DS}. 

The last line in eq. (\ref{Lcomplete})
is the brane Lagrangian (\ref{Lbraneis}), supplemented by 
a higher-order correction with  coefficients $\tau\hat\beta^a$. 
Its role will be discussed below. The identity
\beq
\label{sugraident}
{1\over4}\left[ \hat S {\cal W}^a{\cal W}^a \right]_F
=-2\left[ (\hat S+\ov{\hat S})\Omega^a\right]_D + {\rm derivative}
\eeq
can also be used as a definition of the gauge curvature chiral multiplets 
${\cal W}^a$.

Theory (\ref{Lcomplete}) has a very simple Einstein term since only the
bulk Lagrangian contributes:
\beq
\label{Ein1}
{\cal L}_{\rm Einstein} = 
-{1\over2}eR \left[ (z_0\ov z_0 C_T)^{3/2}(2C)^{-1/2}\right]\,,
\eeq
where $z_0$, $C_T$ and $C$ denote the lowest components of $S_0$,
$V_T$ and $V$. As mentioned already in eq. (\ref{Upsilonis}), the 
Einstein frame is selected by the condition
\beq
\label{Ein2}
\left({z_0\ov z_0 C_T\over 2C}\right)^{-3/2} 
=2\kappa^2 C.
\eeq
The Einstein frame will be used below. 

Since theory (\ref{Lcomplete}) contains `auxiliary multiplets' which 
can be eliminated, we will consider two versions related by 
chiral-linear duality acting on the dilaton multiplet:
\begin{itemize}
\item
The {\it linear version} is obtained by solving for $L_T$, $U$ and $S$.
The dynamical multiplets are then $L$, $T$, $M$, $\Omega^a$ and the 
brane multiplet $\hat L$ or $\hat S$. The dilaton is described by the 
linear multiplet $L$, which also includes the massless component
$G_{4\mu\nu\rho}$ of the four-form field.
\item
The {\it chiral version} is obtained by solving for $L_T$, $U$ and 
$V$, the dynamical multiplets being then $S$, $T$, $M$, $\Omega^a$ 
and the brane multiplet $\hat L$ or $\hat S$. The dilaton is described 
by the real part $\Re S$ of the scalar component of the
chiral multiplet $S$, while $\Im S$ is a component
of the six-form field. 
\end{itemize}

For our purposes, it is useful to simplify the theory by solving for
$L_T$ and $U$. Their field equations respectively imply that 
$V_T= T+\ov T- 2\ov Me^AM$, with a chiral modulus multiplet $T$,
and that the superpotential is a cubic gauge invariant 
function of $M$ which we symbolically write $W(M)=\alpha M^3$,
up to a possible constant (which would break supersymmetry).
The result is the following effective Lagrangian:
\beq\label{Lcomplete2}
\begin{array}{rcl}
\L &=& 
\biggl[-\biggl(S_0\overline{S_0}(T+\ov T- 2\ov Me^AM)\biggr)^{3/2}
(2V)^{-1/2} -(S+\overline S)(V+2\Omega) \crbig
&& 
+V(\epsilon|\alpha M^3|^2-2\delta\overline Me^AM)\biggr]_D
-\tau\left[V(T+\ov T- 2\ov Me^AM)\hat V^2
-{1\over 2}(\hat S+\overline{\hat S})V\hat V\right]_D
\crbig
&& 
+\biggl[S_0^3W(M)
+{1\over4}\sum_a (\beta^aT + \tau\hat\beta^a\hat S)
{\cal W}^a{\cal W}^a \biggr]_F.
\end{array}
\eeq

We first omit the higher-order corrections: $\beta^a=\hat\beta^a=
\delta=\epsilon=0$. All terms in the Lagrangian are then obtained from the
reduction of the higher-dimensional bosonic action (\ref{Sfinal}) and
of the brane action (\ref{brane1}), supplemented by $N=1$ supersymmetry. 
We also choose to describe the brane multiplet by the chiral multiplet
$\hat S$ by solving for $\hat V$. Then, with identity (\ref{sugraident}),
\beq
\begin{array}{rcl}
\L &=& 
\biggl[-\biggl(S_0\overline{S_0}(T+\ov T- 2\ov Me^AM)\biggr)^{3/2}
(2V)^{-1/2} \crbig
&& \displaystyle{
-\biggl( S+\ov S-{\tau\over16}{(\hat S+\overline{\hat S})^2
\over T+\ov T- 2\ov Me^AM} \biggr) V \biggr]_D
+\biggl[S_0^3W(M) +{1\over4} S\sum_a c^a{\cal W}^a{\cal W}^a \biggr]_F, }
\end{array}
\eeq
and solving for $V$ leads to the chiral version, in which the (bulk)
dilaton is described by $S$. It is as usual defined by
\beq
\label{chiralgen}
\L_{\rm chiral} =
-{3\over2}\left[S_0\overline{S_0}\,e^{-K/3}\right]_D
+\Bigl[{1\over4}\sum_a f^a{\cal W}^a{\cal W}^a 
+ S_0^3W(M)\Bigr]_F.
\eeq
The real K\"ahler potential is
\beq\label{Kis1}
K=
-\log\biggl(S+\overline S
-{\tau\over 16}{(\hat S+\overline{\hat S})^2\over T+\overline T-2\ov Me^AM}
\biggr)-3\log \left(T+\overline T -2\ov Me^AM \right), 
\eeq
and the holomorphic gauge kinetic functions are simply
\beq
f^a=c^aS.
\eeq
It is important to realize that the brane kinetic terms affect the
dilaton K\"ahler potential, and that this modification cannot be moved
into the gauge kinetic function by a holomorphic redefinition of the 
chiral $S$: the brane kinetic terms are not harmonic. 

Suppose nevertheless that we insist on defining the dilaton as the 
real quantity
\beq
\varphi = \Re S 
-{\tau\over 32}{(\hat S+\overline{\hat S})^2\over T+\overline T-2\ov MM},
\eeq
as suggested by the K\"ahler potential (\ref{Kis1}).
The coupling for the gauge group factor $G^a$ can then be written 
as\footnote{It is the `wilsonnian gauge coupling'.}
\beq
{1\over g_a^2} = \Re f^a = c^a\left(\varphi 
+{\tau\over 32}{(\hat S+\overline{\hat S})^2\over T+\overline T-2\ov MM}
\right).
\eeq
In this point of view, the brane contribution appears as a correction
to the gauge coupling. However, one cannot find a {\it holomorphic}
function $f^a$ with the field variable $\varphi$
and this choice of dilaton field is not compatible with
the supermultiplet structure required when writing the supergravity 
Lagrangian in the chiral version.

The addition of the higher-order corrections is straightforward. In the
chiral version, the K\"ahler potential becomes
\beq
\label{fullK}
\begin{array}{rcl}
K &=& \displaystyle{
-\log\biggl(S+\overline S
-{\tau\over 16}{(\hat S+\overline{\hat S})^2\over T+\overline T-2\ov Me^AM}
+2\delta\ov Me^AM -\epsilon|\alpha M^3|^2 \biggr) }
\crbig
&&-3\log \left(T+\overline T -2\ov Me^AM \right)
\end{array}
\eeq
while the gauge kinetic functions read
\beq
\label{fullfa}
f^a = c^a S + \beta ^a T + \tau\hat\beta^a \hat S.
\eeq
The `natural' definition of the dilaton suggested by the K\"ahler potential
is now 
\beq
\label{dilaton2}
\varphi = \Re S
-{\tau\over 32}{(\hat S+\overline{\hat S})^2\over T+\overline T-2\ov MM}
+\delta\ov MM -{1\over2}\epsilon|\alpha M^3|^2 ,
\eeq
and in terms of this dilaton, the gauge couplings become
\beq
\label{gaugephi}
\begin{array}{rcl}
\displaystyle{1\over g_a^2} &=&  c^a\varphi + \beta^a \Re T
-c^a\delta\ov MM +{1\over2} c^a\epsilon |\alpha M^3|^2
\crbig
&& +{1\over2}\tau( T+\overline T-2\ov MM)
\left({1\over16}c^a\left({\hat S+\overline{\hat S}\over 
T+\overline T-2\ov MM}\right)^2
+\hat\beta^a {\hat S+\overline{\hat S}\over 
T+\overline T-2\ov MM}\right). 
\end{array}
\eeq

Returning to the Lagrangian (\ref{Lcomplete}), the field equation 
relating $\hat V$ and $\hat S$ is
\beq
\label{hatVeom}
\hat V = {1\over4} {\hat S+\overline{\hat S}\over 
T+\overline T-2\ov Me^AM},
\eeq
and the lowest component $\hat C$ of $\hat V$ has been identified 
with $\tilde X = \kappa^{-1}X$, which is the brane modulus in the
direction $x^4$, in Planck units. The gauge couplings can then finally
be expressed as
\beq
\label{gaugephi2}
\begin{array}{rcl}
\displaystyle{1\over g_a^2} &=&  c^a\varphi + \beta^a \Re T
-c^a\delta\ov MM +{c^a\epsilon\over2}|\alpha M^3|^2
\crbig
&& +{\tau\over2}( T+\overline T-2\ov MM)
\left(c^a \tilde X^2+4\hat\beta^a \tilde X\right). 
\end{array}
\eeq

The linear version is interesting. Solving in eq. (\ref{Lcomplete2}) 
for $S$ implies $V= L-2\Omega$, and the resulting effective supergravity
reads
\beq
\label{linear1}
\begin{array}{rcl}
\L_{\rm linear} &=& 
\biggl[-{1\over\sqrt2}\biggl(S_0\overline{S_0}
(T+\ov T- 2\ov Me^AM)\biggr)^{3/2}
(L-2\Omega)^{-1/2} \crbig
&& 
+(L-2\Omega)\biggl(\epsilon|\alpha M^3|^2-2\delta\overline Me^AM
-\tau (T+\ov T- 2\ov Me^AM)\hat V^2
\crbig
&&+{\tau\over 2}(\hat S+\overline{\hat S})\hat V\biggr)\biggr]_D
+\biggl[S_0^3W(M)
+{1\over4}\sum_a (\beta^aT + \tau\hat\beta^a\hat S)
{\cal W}^a{\cal W}^a \biggr]_F.
\end{array}
\eeq
In this case, with the identification $\hat C=\tilde X$ and with the
field equation (\ref{hatVeom}) relating $\hat S$ and $\hat C$,
computing the gauge couplings leads easily to
\beq
\label{gaugelin}
\begin{array}{rcl}
\displaystyle{1\over g_a^2} &=& 
{1\over2}c^a\biggl({z_0\overline{z_0}(T+\ov T- 2\ov MM)
\over 2C}\biggr)^{3/2}
+{1\over2} c^a\epsilon |\alpha M^3|^2-c^a\delta\overline MM \crbig
&&+{\tau\over2}(T+\ov T- 2\ov MM) [c^a\tilde X^2 + 4\hat\beta^a\tilde X]
+\beta^a\Re T \,.
\end{array}
\eeq
Comparing with expression (\ref{gaugephi2}),
we find that
$$
2\varphi = \biggl({z_0\overline{z_0}(T+\ov T- 2\ov MM)
\over 2C}\biggr)^{3/2}
$$
or, in the Einstein frame, with condition (\ref{Ein2}) and $C_T=T+\ov T
-2\ov MM$, 
\beq
\label{varphiC}
\varphi = {1\over 4\kappa^2 C}.
\eeq
The compatibility of expressions (\ref{gaugelin}) and (\ref{fullfa}) 
follows then from  the field equation of the vector multiplet $V$ 
in theory (\ref{Lcomplete2}), which is chiral-linear duality:
\beq
\label{chilin}
\begin{array}{rcl}
S+\ov S &=& \biggl({S_0\overline{S_0}(T+\ov T- 
2\ov Me^AM)\over 2V}\biggr)^{3/2} 
+\epsilon|\alpha M^3|^2-2\delta\overline Me^AM
\crbig
&&-\tau (T+\ov T- 2\ov Me^AM)\hat V^2
+{\tau\over 2}(\hat S+\overline{\hat S})\hat V \,.
\end{array}
\eeq

To summarize, in the chiral version of the effective supergravity, the 
kinetic Lagrangian of the five-brane modulus introduces a quadratic, 
non-harmonic correction to the dilaton in the K\"ahler potential. The 
holomorphic gauge functions and the wilsonnian gauge couplings are not 
affected by these terms. In the linear version, the kinetic brane Lagrangian 
generates quadratic, non-harmonic corrections to the field-dependent 
wilsonnian gauge couplings. 

The higher-order brane contribution 
$\Delta_{\rm brane}={1\over4}\tau\sum_a\hat\beta^a
[\hat S{\cal W}^a {\cal W}^a]_F$ is similar to the familiar gauge thresholds 
in the modulus $T$, with coefficients $\beta^a$. We have seen that the 
self-duality of the three-index tensor on the brane world-volume leads in 
four dimensions to a chiral-linear duality. In the effective supergravity, 
this duality requires invariance under variations of $\hat S$ by an 
imaginary constant. Then, $D$-terms should depend on $\hat S+\ov{\hat S}$, 
and with our set of multiplets, there is a unique $F$-term compatible 
with this symmetry: the higher-order correction $\Delta_{\rm brane}$ 
\footnote{The authors of refs. \cite{LOW3, CM} failed to recognize 
the importance of
self-duality of the world-volume three-form. They attempted to describe
the brane modulus with a chiral multiplet and introduced a quadratic 
holomorphic $F$-density forbidden by chiral-linear duality and unrelated 
to brane kinetic terms. The resulting supergravity theory is 
incorrect.}.

This second brane contribution is not generated by reduction of the PST brane
action (\ref{brane1}), as $T$-dependent threshold corrections do not follow 
from reduction of the bosonic action (\ref{Sfinal}). In that sense, 
they can be regarded as higher-order terms.

The presence of quadratic and linear brane contributions has been 
established in the background calculation of Lukas, Ovrut and Waldram
\cite{LOW1, LOW, LOW3}. They have in particular computed the gauge 
couplings for a set of branes located at fixed positions along $S^1$. 
These positions correspond to constant background values of our scalar
field $\tilde X$. To compare with our result, it is easier (and sufficient)
to consider a single brane, two gauge couplings and a single modulus
$T$, as in our reduction. The variables used by LOW 
are then the position $z$ along the interval $S^1/\Ztwo$ and 
three charges $\beta^{(0)}$, $\beta^{(2)}$ and $\beta^{({\rm 5b.})}$ 
associated with the two fixed planes and the brane. The variable
$z$ is normalized in the interval $[0,1]$ and the charges are 
quantized: $\beta^{(0)}$ and $\beta^{(2)}$ are half integers, 
$\beta^{({\rm 5b.})}$ is an integer, and the cohomology (or background)
condition implies $\beta^{(0)}+\beta^{(2)}+\beta^{({\rm 5b.})}=0$. The
gauge couplings found by LOW are then
\beq
\label{LOWg2}
\begin{array}{rcl}
{1\over g_1^2} &=&  \Re S +{\epsilon_S\over8\pi} \Re T[\beta^{(0)}+
(1-z)^2\beta^{({\rm 5b.})} ], \crbig
{1\over g_2^2} &=&  \Re S +{\epsilon_S\over8\pi} \Re T[-(\beta^{(0)}
+\beta^{({\rm 5b.})}) + z^2\beta^{({\rm 5b.})}], 
\crbig
{1\over g_1^2}-{1\over g_2^2} &=& {\epsilon_S\over4\pi} \Re T
[\beta^{(0)}+\beta^{({\rm 5b.})} - z\beta^{({\rm 5b.})}] ,
\end{array}
\eeq
with a dimensionless (arbitrary) parameter $\epsilon_S$ related
to the Calabi-Yau volume and the $S^1$ radius. 
Compare these quantities with our expression (\ref{gaugephi2}), 
with $M=0$ and $c^a=1$:
\beq
\label{ourLOW}
\begin{array}{rcl}
{1\over g_{1,2}^2} &=&  \varphi + \Re T[\beta^{1,2} + \tau\tilde X^2
+4\tau\hat\beta^{1,2}\tilde X ], \crbig
{1\over g_1^2}-{1\over g_2^2} &=& \Re T[\beta^1-\beta^2
+4\tau(\hat\beta^1-\hat\beta^2)\tilde X] .
\end{array}
\eeq
Our parameters are not normalized or quantized. If we merely write
$\tilde X=\tilde\lambda z$, both sets of equations coincide with the 
trivial statement 
$\tau={\epsilon_S\over8\pi\tilde\lambda^2}\beta^{({\rm 5b.})}$ 
and the non-trivial relations
\beq
\label{identfinal}
\beta^1=-\beta^2 = {\epsilon_S\over8\pi}(\beta^{(0)}+\beta^{({\rm 5b.})}), 
\qquad\qquad
\hat\beta_1 = -{\tilde\lambda\over2}, \qquad\qquad
\hat\beta_2=0.
\eeq
These equations are predictions obtained from the solution of the 
background condition which specify in parts our four arbitrary 
threshold parameters. They are specific properties of M-theory 
compactified on $O_7$. Our effective supergravity reproduces then nicely
the background found by LOW. Notice in passing 
that the dilaton field is incorrectly identified in eqs. (\ref{LOWg2})
as the real part of the chiral $S$. We have seen that the correct 
identification is $\varphi=(4\kappa^2 C)^{-1}$, in the linear version of the 
theory. A background calculation is not sufficient to reach 
this conclusion: a constant value $z$ of the brane modulus
can be the background value of any kind of 
multiplets (vector, linear, chiral). 

\subsection{The scalar potential}

We close this section by a discussion of the impact of the five-brane
modulus on the supergravity scalar potential. 

We first use the chiral version, defined by the K\"ahler potential
(\ref{Kis1}) and the superpotential $W(M)$. We concentrate
on the potential at $M=0$ \footnote{Since $M$ is a 
charged field, the potential is always stationary at $M=0$.}. We however
assume that the superpotential can be nonzero in this limit: this is the
case if the component $G_{4ijk}$ of the four-form field is a non-zero 
constant breaking supersymmetry. 
As usual, the potential (in the Einstein frame) reads
$$
\kappa^4 \, V(S,T, \hat S) = 
e^K\,W\ov W\,\left[\sum_{IJ}(K_I+W^{-1}W_I)(K^J+\ov W^{-1}\ov W^J)
(K^I_J)^{-1} -3\right] ,
$$
where $K_I={\partial K\over\partial z^I}$, $K^I=(K_I)^*\ldots$, and
$z_I=(S,T,\hat S)$. In the absence of the five-brane field $\hat S$,
the potential takes the simple form 
$\kappa^4 V(S,T) = e^K\,W\ov W$.
An explicit calculation shows that this result {\it is not} affected by 
the contributions of the five-brane modulus. The complete scalar potential 
at $M=0$ in the chiral version of the theory is then:
\beq
\kappa^4\, V (S,T,\hat S)  
= {W\ov W\over 
(S+\ov S-{\tau\over 16}{(\hat S+\ov{\hat S})^2\over T+\ov T})(T+\ov T)^3}\,.
\eeq
This result can be easily understood in the linear version of the theory,
or in the original expression (\ref{Lcomplete2}) of the Lagrangian. 
The five-brane terms do not include any contribution to the 
scalar potential: we have discussed this point in paragraph 3.4. 
In the linear version, the scalar potential is then completely 
independent from the brane modulus $\hat C$. This statement would 
remain true with several five-branes, since each contributes by adding 
to eq. (\ref{Lcomplete2}) a similar term, without any scalar potential.

The appearance of a dependence in $\hat S$ of the potential in the 
chiral version follows from the chiral-linear duality equation 
(\ref{chilin}). The relation between $S$ and $C$ is modified by the 
five-brane to become
$$
S+\ov S - {\tau\over16}{(\hat S+\ov{\hat S})^2\over T+\ov T} 
= {1\over2\kappa^2 C}
$$
with $M=0$ and in the Einstein frame. It is the dependence on $C$ of the
scalar potential in the linear version which induces a dependence on 
$\hat S$ in the chiral version. As a consequence, the five-brane
modulus does not produce a new minimum equation:
$$
{\partial V\over \partial\hat S} = -{\tau\over8}\,
{\hat S+\ov{\hat S}\over T+\ov T}\,{\partial V\over\partial S} 
$$
and $\partial V\over\partial S$ is not zero. The impact of the five-brane
modulus on the effective scalar potential is then a simple redefinition
of the chiral dilaton field $S$ as a function of the (unchanged) 
$C$ of the linear multiplet formulation.

\section{Concluding remarks}\label{secfinal}
\setcounter{equation}{0}

Even if the five-brane is not a perturbative object, it is interesting to 
consider the brane corrections to the four-dimensional effective 
supergravity from the perspective of string perturbation theory. 
The string loop-counting field is our multiplet $V$
with dilaton $C$, and a $n$-loop
term in the Wilson Lagrangian is characterized by a factor 
$V^{(3n-1)/2}$ \cite{CFV}. According to eq. (\ref{Lcomplete}), 
the kinetic Lagrangian of the five-brane modulus multiplet is similar 
to a one-loop correction, linear in $V$. 
The origin of this factor is simple: the kinetic terms are 
normalized by the world-volume induced metric $\sqrt{-\hat g} 
\sim e^{-3\sigma-2\gamma} \sim C C_T^{-2}$. Compare now with the one-loop
corrections in the modulus $T$, which are completely understood in 
compactifications of heterotic strings. Two kinds of contributions arise 
\cite{DKL, DFKZ}. The first is a real gauge-group independent 
term proportional to the
K\"ahler potential $-3\log(T+\ov T)$, the `Green-Schwarz' term. 
The second one is a gauge-group dependent correction which involves 
a holomorphic function. In the chiral version, the Green-Schwarz term 
corrects the K\"ahler potential of the $S$ field, it can be regarded as 
a wave-function renormalization of this field. The second term is then a
correction to the gauge kinetic functions $f^a$. The similarity with
the five-brane contributions in the Lagrangian (\ref{Lcomplete})
is obvious. In the case of the volume modulus $T$, 
the one-loop corrections can be 
understood in terms of a cancellation of target-space duality anomalies. 
The analogy suggests that anomalies could also 
help to understand the structure of our five-brane contributions
\cite{BDS2}. 

\vspace{.5cm}
\begin{center}{\bf Acknowledgements}\end{center}
\noindent
The authors have benefitted from discussions with A. Bilal and C. 
Kounnas. This research was supported by the European 
Union under the TMR contract ERBFMRX-CT96-0045, the Swiss National 
Science Foundation and the Swiss Office for Education and Science.
\vspace{.5cm}

\section*{Appendix: Notations and conventions}
\label{appnotation}
\renewcommand{\theequation}{A.\arabic{equation}}
\setcounter{equation}{0}

{\it Coordinates and metrics:}

\noindent
For coordinates, our notation is: \\

\begin{tabular} {lll}
$D=11$ curved space-time: & $x^M$ & $M=0,\ldots,10$ \\
$D=4$ curved space-time: & $x^\mu$ & $\mu =0,1,2,3$ \\
$S^1/\Ztwo$ direction: & $x^4$ & \\
Calabi-Yau directions, real: & $x^a$ & $a=5,\ldots,10$  \\
Calabi-Yau complex (K\"ahler) coordinates: $\qquad$& 
$z^i$, $\ov z^{\ov i}\qquad$ & $i=1,2,3$ \\
Five-brane world-volume coordinates: & $y^{\hat m}$ &
$\hat m = \hat 0,\ldots,\hat 5$
\end{tabular} \\[2mm]

\noindent
For reduction purposes, we simply use
$$
z^l = {1 \over \sqrt{2}}\left(x^l+ix^{l+3}\right)\,,
\qquad\qquad 
\ov z^{\ov l} = {1 \over \sqrt{2}}\left(x^l-ix^{l+3}\right)\,,
\qquad\qquad l=1,2,3.
$$ 
$\epsilon_{ijk}$ is the $SU(3)$--invariant Calabi-Yau tensor 
with $\epsilon_{123}=\epsilon_{\ov{123}}=1$.

\noindent
The space-time metric has signature $(-,+,+,\ldots,+).$ 
The reduction of the eleven-dimensional metric is defined by
\beq\label{elemetric}
g_{MN}=\pmatrix{
e^{-\gamma-2\sigma}g_{\mu\nu} & 0 & 0 \cr
0 & e^{2\gamma-2\sigma} & 0 \cr
0 & 0 & e^\sigma\delta_{i\ov j} \cr
}.
\eeq

\medskip
\noindent{\it Antisymmetric tensors:}

\noindent
Antisymmetrization of $n$ indices has unit weight:
$$
A_{[M_1\ldots M_n]} = {1\over n!}\Bigl(A_{M_1\ldots M_n} \pm 
(n!-1){\rm \,\,permutations}\Bigr).
$$

\medskip
\noindent{\it Differential forms:}

\noindent
For a $p$--index antisymmetric tensor, we define
$$
A_p = {1\over p!}A_{M_1\ldots M_p}\, 
dx^{M_1}\wedge\ldots\wedge dx^{M_p}. 
$$
Then, 
$$
\begin{array}{l}
A_p\wedge B_q = {1\over p!q!}A_{M_1\ldots M_p}
B_{M_{p+1}\ldots M_{p+q}} \,dx^{M_1}\wedge\ldots\wedge 
dx^{M_{p+q}} = C_{p+q}, \crbig
C_{M_1\ldots M_{p+q}} = {(p+q)!\over p!q!}A_{[M_1\ldots M_p} 
B_{M_{p+1}\ldots M_{p+q}]}.
\end{array}
$$
The exterior derivative is $d=\partial_M\,dx^M$. 
The curl $F_{p+1} = dA_p$ of a $p$-form reads then
$$
\begin{array}{rcl}
dA_p &=& {1\over p!} (\partial_MA_{N_1\ldots N_p})\,
dx^M\wedge dx^{N_1}\wedge\ldots\wedge dx^{N_p}  \crbig
&=& {1\over (p+1)!} F_{M_1\ldots M_{p+1}} \, dx^{M_1}\wedge\ldots
\wedge dx^{M_{p+1}}, \crbig
F_{M_1\ldots M_{p+1}} &=& (p+1)\, \partial_{[M_1}A_{M_2\ldots 
M_{p+1}]} \crbig
&=& \partial_{M_1}A_{M_2\ldots M_{p+1}} \pm \,\, p
{\rm \,\,\, cyclic\,\,permutations}\,.
\end{array}
$$
The volume form in $D$ space-time dimensions is
$dx^{M_1}\wedge\ldots\wedge dx^{M_D} = \epsilon^{M_1\ldots 
M_D}\,d^Dx$.
\\ \noindent We use analogous conventions for forms in four
space-time dimensions.

\end{document}